\documentclass[10pt,journal,compsoc]{IEEEtran}
\usepackage{spreadtab}
\usepackage{amsmath,amssymb,amsfonts}
\usepackage{algorithmic}
\usepackage{graphicx}
\usepackage{textcomp}
\usepackage[T1]{fontenc}
\usepackage{graphicx}
\usepackage{pgfplots}
\pgfplotsset{every tick label/.append style={font=\normalsize}}
\usepackage{booktabs}
\usepackage{multirow}
\usepackage{multicol}
\usepackage{hhline}
\usepackage{subcaption}
\usepackage{eqnarray}
\usepackage[ruled, noend, noline]{algorithm2e}
\usepackage{setspace}
\usepackage{xcolor}
\usepackage{mathtools, nccmath}
\DeclarePairedDelimiter{\nint}\lfloor\rceil

\ifCLASSOPTIONcompsoc
  \usepackage[nocompress]{cite}
\else
  \usepackage{cite}
\fi
\ifCLASSINFOpdf
\else
\fi
\begin{document}
%
\title{Performance Modeling of Epidemic Routing in Mobile Social Networks with Emphasis on Scalability}
%
%
%
%

\author{Leila~Rashidi,
        Amir~Dalili-Yazdi,
        Reza~Entezari-Maleki,
        Leonel~Sousa,~\IEEEmembership{Senior~Member,~IEEE}
        and~Ali~Movaghar,~\IEEEmembership{Senior~Member,~IEEE}
\IEEEcompsocitemizethanks{\IEEEcompsocthanksitem L. Rashidi is with the Department of Computer Science, University of Calgary, Calgary,
AB, Canada.\protect\\
E-mail: leila.rashidi@ucalgary.ca
\IEEEcompsocthanksitem A. Dalili-Yazdi and A. Movaghar are with the Department of Computer Engineering, Sharif University of Technology, Tehran, Iran.\protect\\
E-mail: dalili@ce.sharif.edu; movaghar@sharif.edu
\IEEEcompsocthanksitem R. Entezari-Maleki is with the School of Computer Engineering, Iran University of Science and Technology, Tehran, Iran and INESC-ID, Instituto Superior T\'{e}cnico, Universidade de Lisboa, Lisbon, Portugal.\protect\\
E-mail: entezari@iust.ac.ir
\IEEEcompsocthanksitem L. Sousa is with INESC-ID, Instituto Superior T\'{e}cnico, Universidade de Lisboa, Lisbon, Portugal.\protect\\
E-mail: las@inesc-id.pt}
\thanks{This paper is an extended version of the conference paper \cite{8843121} published in the IEEE 27th International Symposium on Modeling, Analysis, and Simulation of Computer and Telecommunication Systems (MASCOTS), 2019, pages: 201-213.}}

\IEEEtitleabstractindextext{%
\begin{abstract}
This paper investigates the performance of epidemic routing in mobile social networks. It first analyzes the time taken for a node to meet the first node of a set of nodes restricted to move in a specific subarea. Afterwards, a monolithic Stochastic Reward Net (SRN) is proposed to evaluate the delivery delay and the average number of transmissions under epidemic routing by considering skewed location visiting preferences. This model is not scalable enough, in terms of the number of nodes and frequently visited locations. In order to achieve higher scalability, the folding technique is applied to the monolithic model, and an approximate folded SRN is proposed to evaluate performance of epidemic routing. Discrete-event simulation is used to validate the proposed models. Both SRN models show high accuracy in predicting the performance of epidemic routing. We also propose an Ordinary Differential Equation (ODE) model for epidemic routing and compare it with the folded model. Obtained results show that the folded model is more accurate than the ODE model. Moreover, it is proved that the number of transmissions by the time of delivery follows uniform distribution, in a general class of networks, where positions of nodes are always independent and identically distributed.
\end{abstract}

}

\maketitle

\IEEEdisplaynontitleabstractindextext

%
\IEEEpeerreviewmaketitle

\IEEEraisesectionheading{\section{Introduction}\label{sec:introduction}}

%
%
%
%
\IEEEPARstart{M}{obile} Social Networks (MSNs) are a kind of Delay Tolerant Networks (DTNs) \cite{cao2013routing} consisting of some mobile nodes that share information with each other using short-range communication technologies \cite{xiao2015home}. Short-range wireless technologies of portable devices, such as smart phones, tablets, and sensors in vehicles, can be used by mobile users to share multimedia, data large-size files, etc. \cite{xiao2014community}. MSNs can be used for opportunistic mobile data offloading and for providing communication during disasters \cite{7961189, 8419321}.
One of the main characteristics of MSNs is that nodes have skewed location visiting preferences \cite{4215676}. In real world scenarios, people visit locations with different frequencies. As an example, every employee visits her/his work place each business day while she/he might prefer to go to a shopping center only once a week. Specifically, we tend to spend most of our time at a few frequently visited locations \cite{4215676, 7857091}. We call such a location \textit{community}. 

Despite various network models considered in the literature to analyze the performance of routing in DTNs, the performance of networks where nodes have skewed location visiting preferences has not been well-studied. Scalability is one of the most important challenges in performance analysis of heterogeneous networks. In this paper, we focus on evaluating the performance of epidemic routing \cite{vahdat2000epidemic}, in a scalable way, considering skewed location visiting preferences for nodes in a heterogeneous network. The aim is to compute the average and Cumulative Distribution Function (CDF) of the delivery delay of a message, from the source to the destination, and the average number of transmissions of the message by time of delivery as in \cite{zhang2007performance}. Epidemic routing has the minimum delivery delay and the maximum communication cost in terms of the number of transmissions. Evaluating the average delivery delay and the average number of transmissions of epidemic routing thus provides good insights into the design of efficient routing schemes or network configuration. For instance, it shows the extent to which the delivery of a message could be fast.

We study the first meeting time of a node with a set of nodes moving in a specific part of the area where that node moves. It is worth mentioning that \textit{meeting of a node with a set of nodes} refers the first meeting of a node with a node of the considered set. Characterizing such a meeting time is useful in the performance analysis of MSNs/DTNs, since in various real scenarios some nodes move only in a specific place during a period of time while some nodes move in a larger area freely. Afterwards, we propose a monolithic Stochastic Reward Net (SRN) \cite{muppala1991composite} model to evaluate the performance of epidemic routing in a network where nodes move in a large area, including some communities frequently visited by nodes.
Although the monolithic SRN model is able to evaluate the performance of small networks, it faces the problem of state space explosion when the network scales up in terms of the number of nodes and communities. In order to solve this problem, an approximate SRN model, by applying the folding technique, is proposed. Numerical results show that the number of states in the underlying Markov chain of the folded model is significantly less than that of the monolithic model. The results of both monolithic and folded SRN models are validated by discrete-event simulation. The analytical and simulation results indicate that both monolithic and folded models are accurate enough to evaluate the performance of the epidemic routing in the target networks. In order to prove the superiority of the proposed folding-based approach in evaluation of the performance of large-scale networks, we apply the Ordinary Differential Equation (ODE) approach \cite{zhang2007performance} to model the target network, and then show that the proposed folded model is more accurate than the ODE-based model.

The main contributions of this paper are as follows.
\begin{itemize}
    \item It is demonstrated that the first meeting time of a node moving in an area with a set of nodes moving in a specific subarea is exponentially distributed.
    \item A monolithic SRN model is proposed to evaluate the average and CDF of the delivery delay and the average number of transmissions of epidemic routing in a network consisting of some communities frequently visited by nodes.
    
    \item By applying the folding technique to the proposed monolithic SRN, a scalable approximate SRN is proposed to evaluate the performance of large-scale networks.
    
    \item The validation is done by simulation, comparing the results of both monolithic and folded SRN models. This comparison indicates that the proposed models have a good accuracy.
    
    \item According to both analytical and simulation results, the average number of transmissions is very close to the half of the number of nodes. In order to justify this observation, it is proved that the average number of transmissions is equal to the half of the number of nodes in any network, not only our target network, where positions of nodes are always independent and identically distributed (i.i.d.).
    
    \item In order to compare the proposed folded model with the ODE approach, this approach is also applied to model epidemic routing in the target networks. Comparison of the results of the folded SRN and ODE models with the results obtained from simulation indicates the superiority of the results obtained from the folded model.
\end{itemize}

The rest of this paper is organized as follows. The related state-of-the-art and the main differences to the work presented in this paper are introduced in Section~\ref{sec:relWork}. Section~\ref{sec:networkModel} introduces the target network model and the assumptions made herein. Afterwards, in Section \ref{sec:meetingTime}, we analyze the time it takes a node to meet the first node belonging to a set under specified conditions on the mobility of nodes. A monolithic SRN and an approximate folded SRN are proposed for epidemic routing in the target network model, in Sections~\ref{sec:mono} and \ref{sec:folded}, respectively. Section~\ref{sec:measures} is dedicated to the figures of merit and how to compute them applying the proposed models. Numerical results obtained from the proposed models and by simulation are provided in Section~\ref{sec:perf}. The proposed SRN models are compared in terms of the scalability in Section~\ref{sec:scalability}. Finally, Section~\ref{sec:conc} concludes the paper and provides some directions for future work.

\section{related work}\label{sec:relWork}
In \cite{groenevelt2005message}, it has been shown that the inter-meeting time of two nodes moving in a square is exponentially distributed. 
In~\cite{IBRAHIM2007933}, it has been demonstrated that the time taken for a mobile node to meet a stationary node is exponentially distributed. In this paper that research is pushed forward by studying the distribution function of the time taken for a mobile node, freely moving in an area, to meet one of the nodes moving in a specific subarea.

In \cite{zhang2007performance}, an ODE-based framework has been proposed to evaluate the performance of epidemic routing and its variations. The network considered in \cite{zhang2007performance} consists of a set of nodes moving in a closed area according to a common mobility model, such as random direction or random waypoint models. In \cite{zhang2007performance}, closed-form expressions for some performance measures, such as the average number of transmissions by the time of delivery, were derived using the analytical solution of the proposed ODE model. However, the ODE approach provides limits to the Markov models when the number of nodes tends to infinity~\cite{zhang2007performance}. Thus, it is not accurate to study the performance of networks with a moderate number of nodes~\cite{yang2016delay}. In particular, the average number of infected nodes at the time of delivery, including the destination node, was estimated to be half of the number of nodes in \cite{zhang2007performance}.
In this paper, we model the epidemic routing in a more realistic network model, considering the skewed location visiting preferences. Moreover, we prove that the average number of transmissions by time of delivery is equal to half of the number of nodes, for a general class of networks, where positions of nodes at any time are independent and follow the same Probability Density Function (PDF). This class includes the network considered in \cite{zhang2007performance}, and the exact expression for the average number of transmissions, derived herein, is close to the approximate expression derived in \cite{zhang2007performance}, given that there is only one initial infected node in the network under-study in \cite{zhang2007performance}.

In \cite{ip2008performance}, a network consisting of two classes of nodes has been considered, wherein the inter-meeting time of any two nodes is exponentially distributed. Subsequently, three rates were defined, one per each class as the meeting rate of any two nodes belonging to that class and another as the meeting rate of any two nodes belonging to different classes. Afterwards, epidemic routing was modeled as a Continuous Time Markov Chain (CTMC), and then two ODE models were proposed in order to evaluate the performance of large-scale networks. One ODE model is an extension of the model proposed in~\cite{zhang2007performance}, while the other ODE model exploits the Kolmogorov forward equation. The network studied in~\cite{spyropoulos2009routing} is similar to~\cite{ip2008performance}, but an arbitrary number of classes was considered in~\cite{spyropoulos2009routing}. In order to evaluate the performance of epidemic routing and some variants of spray and wait routing, a framework that applies ODE model was proposed in~\cite{spyropoulos2009routing}.

In~\cite{sermpezis2016delay},
asymptotic results and closed-form approximations have been derived for epidemic spreading, considering a contact network with probabilistic meeting rates. Unlike \cite{groenevelt2005message} and \cite{ip2008performance, spyropoulos2009routing, picu2012forecasting, picu2015dtn, sermpezis2016delay}, the models proposed in this paper are not based on fixed meeting rates/probabilities or probabilistic meeting rates; two different movement modes are considered, and the meeting rate of any two nodes changes as the movement mode of at least one of them changes.
In \cite{4215676} and \cite{Hsu:2009:MST:1665838.1665854}, a time-variant community mobility model has been proposed. The ODE model proposed in \cite{zhang2007performance} was extended in \cite{Hsu:2009:MST:1665838.1665854} to evaluate the performance of epidemic routing on a network consisting of two communities. As shown in \cite{Hsu:2009:MST:1665838.1665854}, although the average number of infected nodes as a function of time, obtained from the ODE model, follows a trend similar to that observed in simulation results, the ODE model does not yield a good accuracy.

The network studied in \cite{chaintreau2009age} is similar to \cite{spyropoulos2009routing}, but nodes can move between classes with specific rates. In \cite{chaintreau2009age}, meeting times of all pairs of nodes are assumed to be independent from each other. In~\cite{5719289}, an edge-Markovian dynamic graph model has been proposed for epidemic routing. In that model, the states of the edges change independently from each other. However, in real scenarios, the meeting times of some pairs of nodes depend on each other. This dependency was not considered in \cite{chaintreau2009age} and \cite{5719289}, but it is important to take it into account when studying MSNs, as it is the purpose of this paper.
A family of restricted epidemic routings has been modeled in~\cite{wang2015restricted} by applying Discrete Time Markov Chains (DTMCs). Those models are not scalable, and the number of states exceedingly grows when the number of nodes/communities increases. Moreover, considering slotted time is a shortcoming of the models proposed in \cite{5719289} and \cite{wang2015restricted}, while SRNs are based on continuous time which is more realistic. In~\cite{8681155}, two monolithic and folded SRNs have been proposed for the epidemic content retrieval scheme in DTNs with restricted mobility. In~\cite{wang2015restricted} and \cite{8681155}, each node is assumed to move only within the community to which it belongs while in the networks targeted herein, nodes can freely move in a common area and enter all communities. In~\cite{lee2013forwarding}, the delivery delay under both multi-copy two-hop forwarding and direct forwarding has been studied.  


In~\cite{xiao2015home}, a routing scheme has been proposed for a MSN. Specifically, a 2-D grid was considered on which mobile nodes walk randomly and independently from each other. Each node frequently visits few cells, called \textit{homes}, whereas other cells are less frequently visited. The optimality of the routing scheme was studied in \cite{xiao2015home}, assuming that the inter-meeting time of any two nodes and the time between two consecutive visits of a node to its home are exponentially distributed. Based on these assumptions, the proposed routing scheme was modeled by a CTMC. In \cite{xiao2015home}, the next location of each node is randomly selected from the set of its homes or the set of other cells independently from the current location of that node, and the path a node should traverse to reach the next location was ignored. However, we consider that nodes move according to the random direction mobility model
both when they are in communities and outside of communities. In~\cite{xiao2014community}, the single-copy routing problem was studied considering a MSN with a certain number of locations and slotted time. Three assumptions were made in \cite{xiao2014community}: {\it i}) the time taken for each node to reach a frequently visited location follows an exponential distribution; {\it ii}) there is a throwbox at each frequently visited location; {\it iii}) nodes cannot transfer the message to each other when they are outside of the frequently visited locations. The existence of a throwbox at each frequently visited location and transmission only at frequently visited locations are oversimplifications. The network model adopted herein is more realistic than \cite{xiao2015home} and \cite{xiao2014community}, where the location of a node is considered as a discrete quantity. Also it is assumed that time is continuous unlike~\cite{xiao2014community}.    



\section{Network Model}\label{sec:networkModel}

\begin{figure}
    \centering
   \resizebox{0.3\columnwidth}{!}{
    \includegraphics[]{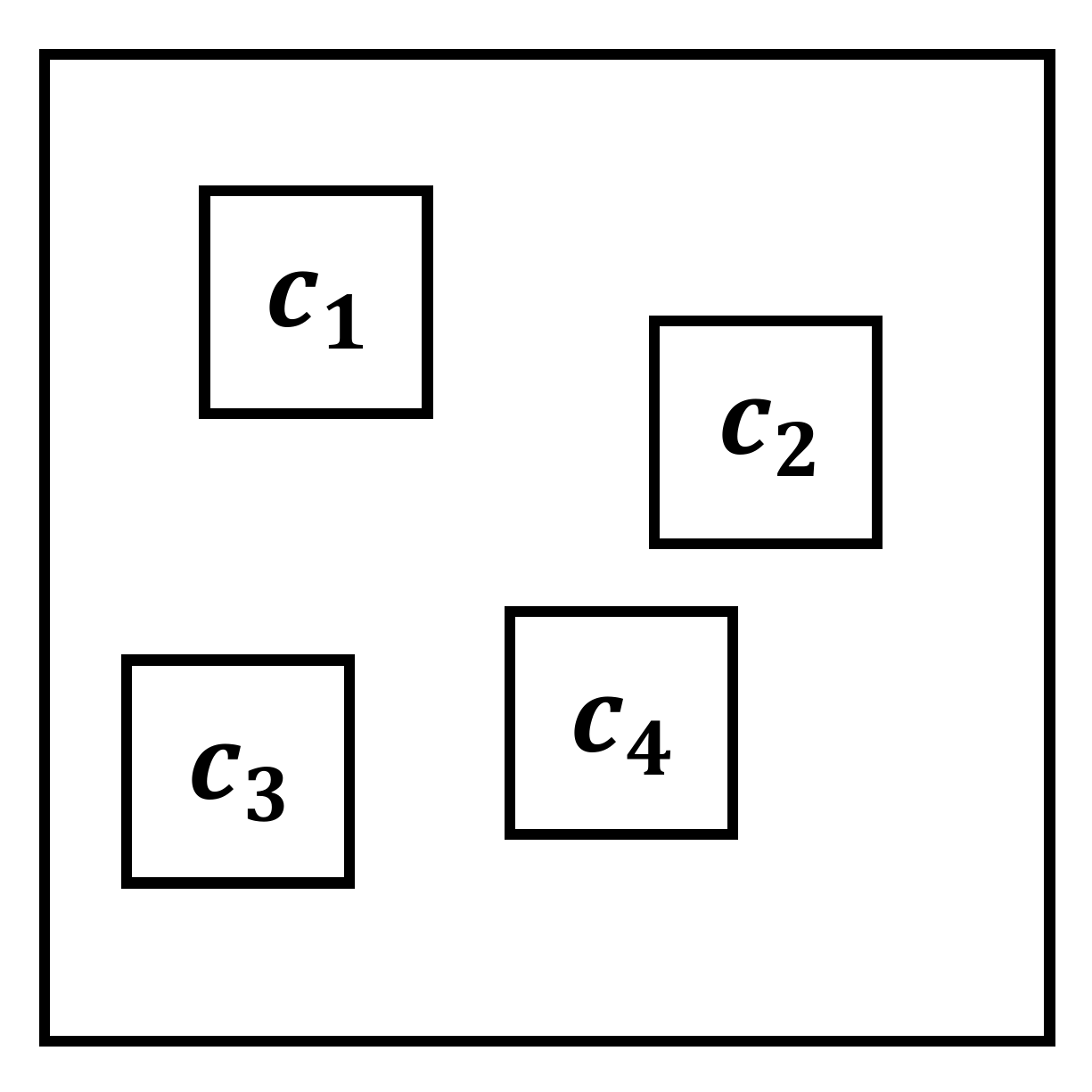}}
    \caption{A network with four communities.}
    \label{fig:example}
\end{figure}

The mobility model considered in this paper is similar to the model proposed in \cite{4215676}, which matches with real-life traces from several scenarios. The network consists of $N$ \mbox{$L_c\times L_c$} communities, denoted by $c_1$, $c_2$, \dots, $c_N$ ($N>1$), located in an $L\times L$ square, called \textit{common area}. For instance, a university campus and each department located in that campus can be considered as common area and a community, respectively. As an example, a network with four communities ($N=4$) is represented in Fig.~\ref{fig:example}. $M$ nodes move in common area such that they visit the communities frequently. In contrast to the network model considered in \cite{8843121}, nodes do not visit a specific community frequently, rather they visit all communities frequently with different frequencies. Initially, nodes are randomly placed within common area with a uniform distribution. 

The communication range of all nodes is fixed, and it is denoted by $R$. We assume each node can move in two different modes: \textit{local} and \textit{roaming}. A node moves within a community or the common area when it is in local and roaming modes, respectively.
In each of these modes, a node moves according to the random direction mobility model, with reflection when hitting boundaries \cite{nain2005properties}. This is more realistic than the torus boundaries. The speed of a node is chosen from $\big(v_{min}, v_{max}\big]$ according to a uniform distribution. The time it takes for each travel in local and roaming modes is distributed exponentially with rates $\alpha$ and $\beta$, respectively. When the movement mode of a node is local and its travel ends, that node changes its movement mode to roaming with probability $P_r$. Moreover, if travel of a node ends while in roaming movement mode, it decides to change its movement mode to local with probability~$P_l$. In case of changing the movement mode to local, the node selects community $c_i$ to move into during local mode with probability~$P_{sel\_i}$. Consider a roaming node that chooses local mode and a community to move in. If it has just ended its travel somewhere in the selected community, the mode is immediately changed; otherwise, it chooses a random position in the selected community and begins to move towards that position by the shortest straight path \cite{Hsu:2009:MST:1665838.1665854}. We call this movement \textit{transitional travel}. Unlike \cite{Hsu:2009:MST:1665838.1665854}, it is assumed that in a transitional travel, a node moves with high speed, denoted by $v_{trans}$, to reach the community soon. This change is applied to the mobility model introduced in~\cite{Hsu:2009:MST:1665838.1665854} in order to make the mobility model theoretically more tractable. Once a node reaches the previously chosen random position in the community, it begins to move in local mode. The notations introduced in this section are summarized in Table~\ref{tab:notationTable}.
\begin{table}
\small
	\renewcommand{\arraystretch}{1}
	\setlength{\tabcolsep}{0.5pt}
	\begin{center}
	\caption{Notations adopted to define the network model}
	\label{tab:notationTable}
	\end{center}
	\centering
	\begin{tabular}{c|c}
		\hline
		\textbf{Notation} & 
		 \textbf{Description}
	
		\\
		
		\hline
		 \hline
			 $N$ &  Number of communities\\
	     \hline
	        $M$ &  Total number of nodes \\
	       \hline
	       $L$ & Edge length of common area \\
	       \hline
	       $L_c$ &  Edge length of each community \\
	       \hline
	       $R$ &  Communication range of each node \\
	       \hline
	       $\alpha$ &  Rate of the duration of a travel in local mode \\
	       \hline
	       $\beta$ &  Rate of the duration of a travel in roaming mode \\
	       \hline
	       $P_r$ &  Probability of changing local mode to roaming mode \\
	       \hline
	       $P_l$ &  Probability of changing roaming mode to local mode \\
	       \hline
	       \multirow{2}{*}{$P_{sel\_i}$} &  Probability of selecting community $c_i$ while  \\
	      \hhline{~~}
	      &   changing the movement mode to local \\
	       \hline
	        $v_{min}$ &  Minimum speed in local/roaming mode  \\
	       \hline
	        $v_{max}$ &  Maximum speed in local/roaming mode  \\
	       \hline
	       $v_{trans}$ &  Speed in a transitional travel  \\
	       \hline
	\end{tabular}
\end{table}

There are two specific nodes called \textit{source} and \textit{destination}. The source wishes to send a message using epidemic routing to the destination. Adopting the terminology from the field of Epidemiology as \cite{spyropoulos2009routing}, the nodes that have (have not) already received the message are called \textit{infected} (\textit{susceptible}). Moreover, \textit{roaming node} and \textit{local node} are used to refer to the nodes that move in roaming and local modes, respectively.
In order to be able to use the benefits of analytical models for analyzing the network, the following assumptions are made, most of them come from the previous works in this area.
\begin{enumerate}
    \item Communities, frequently visited locations, do not overlap each other \cite{xiao2014community, xiao2015home, wang2015restricted}.
    \item Initially, the movement mode of all nodes is roaming. 
     \item Speed $v_{trans}$ is high, the duration of a transitional travel is very short. Based on this assumption, transitional travels are neglected in the proposed models.
    \item The communication range of nodes, $R$, is much less than both the length of the edges of the communities and the common area, $R\ll L_c$ and $R\ll L$. This assumption is very common in the literature \cite{wang2015restricted, groenevelt2005message, 8681155}. 
    \item The first meeting time of any two nodes moving in the same fixed movement mode,
    starting from a random time, is exponentially distributed. This assumption is reasonable when $R\ll L_c$ and $R\ll L$ \cite{groenevelt2005message}, and has been extensively used in recent years \cite{wang2015restricted, 7063247, hsu2016enhanced, yang2016delay, lu2016distance, 8681155}.
    \item The first meeting time of a node constantly moving in roaming mode with a set of nodes constantly moving in the same community in local mode is exponentially distributed. In Section~\ref{sec:meetingTime}, we further explain this kind of first meeting time, and analyze its distribution function. Results obtained from the analysis performed in Section~\ref{sec:meetingTime} justify this assumption.
    \item The delay of a message transmission, which corresponds to a short time, is negligible \cite{groenevelt2005message, zhang2007performance, Kong2008, Zhao:2011:FRN:2030613.2030651, 6151313, peres2013mobile, picu2015dtn, wang2015restricted, 7473919, 8681155, rashidi2020performance}.
\end{enumerate}  

\section{first meeting time of a roaming node with a set of local nodes} 
\label{sec:meetingTime} 
In this section, we analyze the time it takes a node constantly moving in roaming mode, denoted by $A_r$, to meet the first node of a set of local nodes, denoted by $S_l$, which move in the same community. Note that there is no order between the members of the set $S_l$, so the first node of the set indicates the first node which the roaming node meets.

\begin{figure}
    \centering
   \resizebox{0.23\columnwidth}{!}{
    \includegraphics[]{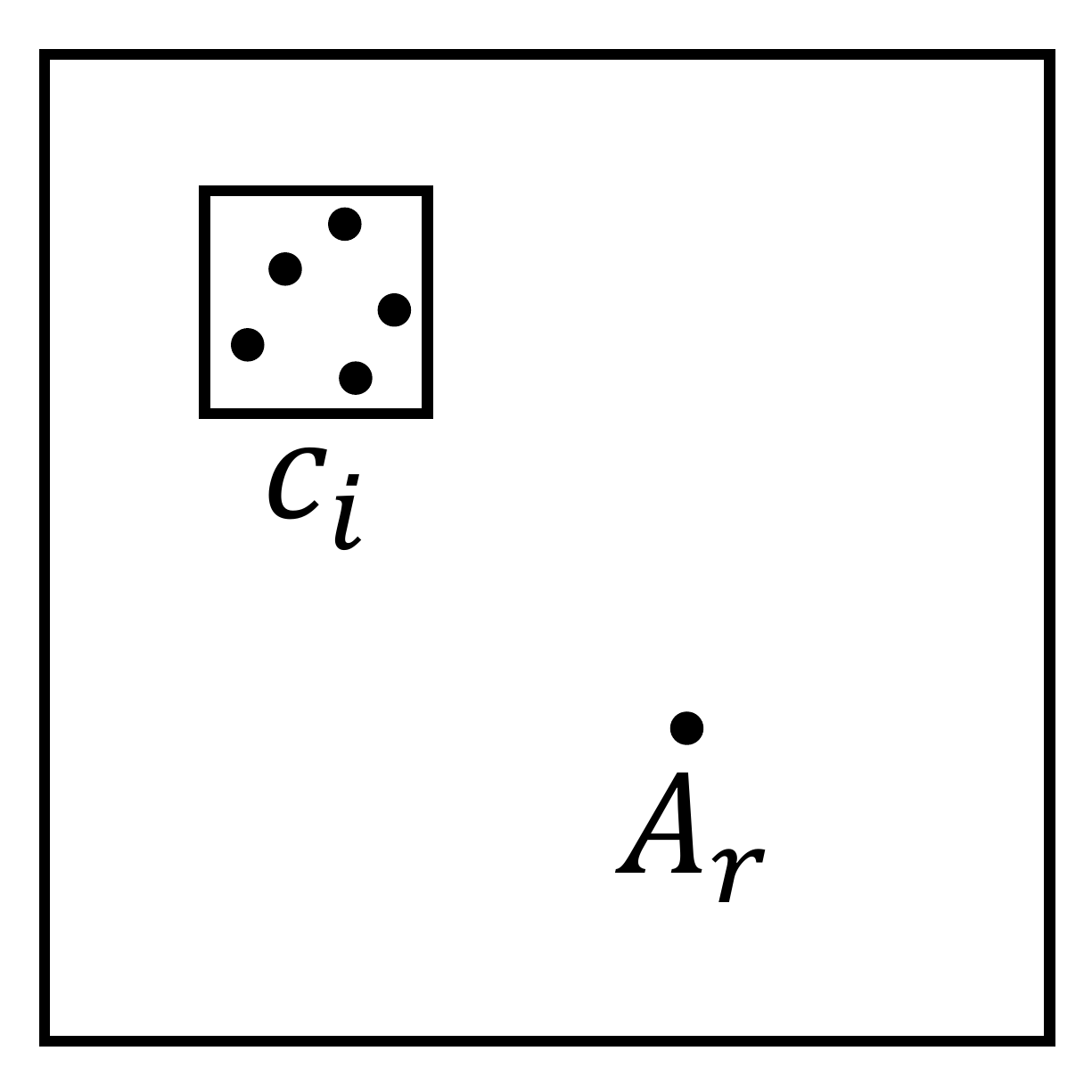}}
    \caption{The situation of node $A_r$ and members of $S_l$ that are in local mode and move in community $c_i$.}
    \label{fig:nodeAndSet}
\end{figure}

For example, Fig.~\ref{fig:nodeAndSet} represents node $A_r$ and five local nodes in community $c_i$ which are members of $S_l$. We show that the time it takes $A_r$ to meet the first node of $S_l$ follows the exponential distribution if $A_r$ and nodes of $S_l$ are initially placed in random locations of the common area and the community, respectively, with a uniform distribution. To this end, a discrete-event simulation (programmed in Java) is conducted. In each simulation run, the initial positions of $A_r$ and nodes of $S_l$ are randomly chosen with a uniform distribution from the common area and the community, respectively, such that $A_r$ is not in communication range of any node belonging to $S_l$. Afterwards, the nodes are moved step by step until $A_r$ meets a node of $S_l$, recording the meeting time in each simulation run. Finally, the CDF of the meeting time is found using the records obtained from the simulation. It is concluded that the obtained CDF exhibits exponential behavior, by using the curve fitting toolbox of Matlab.

\begin{figure}
\centering
\mbox{
\begin{subfigure}[b]{0.25\textwidth}
 \centering
  \begin{tikzpicture}[scale=0.45]
    \begin{axis}[
    xlabel={\Large Time, $t$ ($s$)},
    ylabel={\Large CDF},
    xmin=0, xmax=20000,
    ymin=0, ymax=1,
    xtick={0,4000,8000,12000,16000,20000},
    ytick={0,0.2,0.4,0.6,0.8,1},
    legend pos=south east,
    ymajorgrids=true,
    grid style=dashed,
     every axis plot/.append style={ultra thick},
     every tick label/.append style={font=\Large}
    ]

    \addplot[
    color=blue,
    mark=square,
    mark options={solid},
    ] table [x=time, y=anal, col sep=space] {cdf_1node.txt};\addlegendentry{\Large $1-e^{-\beta\cdot t}$}
 
  \addplot[
    color=red,
    mark=diamond,
    mark options={solid},
    style=dashed
    ] table [x=time, y=simul, col sep=space] {cdf_1node.txt};\addlegendentry{\Large Simulation}

    \end{axis}
\end{tikzpicture}\caption{ $|S_l|=1$}
\end{subfigure}
\hspace{-0.8em}
\begin{subfigure}[b]{0.25\textwidth}
\centering
   \begin{tikzpicture}[scale=0.45]
    \begin{axis}[
    xlabel={\Large Time, $t$ ($s$)},
    ylabel={\Large CDF},
    xmin=0, xmax=6000,
    ymin=0, ymax=1,
    xtick={0,1500,3000,4500,6000},
    ytick={0,0.2,0.4,0.6,0.8,1},
    legend pos=south east,
    ymajorgrids=true,
    grid style=dashed,
     every axis plot/.append style={ultra thick},
     every tick label/.append style={font=\Large}
    ]
    
    \addplot[
    color=blue,
    mark=square,
    mark options={solid},
    ] table [x=time, y=anal, col sep=space] {cdf_4nodes.txt};\addlegendentry{\Large $1-e^{-\beta\cdot t}$}
    
    \addplot[
    color=red,
    mark=diamond,
    mark options={solid},
    style=dashed
    ] table [x=time, y=simul, col sep=space] {cdf_4nodes.txt};\addlegendentry{\Large Simulation}

    \end{axis}
\end{tikzpicture}\caption{ $|S_l|=4$}
\end{subfigure}
}
\vspace*{0.05in}
\\
\mbox{
\begin{subfigure}[b]{0.25\textwidth}
 \centering
     \begin{tikzpicture}[scale=0.45]
    \begin{axis}[
    xlabel={\Large Time, $t$ ($s$)},
    ylabel={\Large CDF},
    xmin=0, xmax=6000,
    ymin=0, ymax=1,
    xtick={0,1500,3000,4500,6000},
    ytick={0,0.2,0.4,0.6,0.8,1},
    legend pos=south east,
    ymajorgrids=true,
    grid style=dashed,
     every axis plot/.append style={ultra thick},
     every tick label/.append style={font=\Large}
    ]

    \addplot[
    color=blue,
    mark=square,
    mark options={solid},
    ] table [x=time, y=anal, col sep=space] {cdf_7nodes.txt};\addlegendentry{\Large $1-e^{-\beta\cdot t}$}
 
    \addplot[
    color=red,
    mark=diamond,
    mark options={solid},
    style=dashed
    ] table [x=time, y=simul, col sep=space] {cdf_7nodes.txt};\addlegendentry{\Large Simulation}

    \end{axis}
\end{tikzpicture}\caption{ $|S_l|=7$}
\end{subfigure}
\hspace{-0.8em}
\begin{subfigure}[b]{0.25\textwidth}
\centering
      \begin{tikzpicture}[scale=0.45]
    \begin{axis}[
    xlabel={\Large Time, $t$ ($s$)},
    ylabel={\Large CDF},
    xmin=0, xmax=6000,
    ymin=0, ymax=1,
    xtick={0,1500,3000,4500,6000},
    ytick={0,0.2,0.4,0.6,0.8,1},
    legend pos=south east,
    ymajorgrids=true,
    grid style=dashed,
     every axis plot/.append style={ultra thick},
     every tick label/.append style={font=\Large}
    ]

    \addplot[
    color=blue,
    mark=square,
    mark options={solid},
    ] table [x=time, y=anal, col sep=space] {cdf_10nodes.txt};\addlegendentry{\Large $1-e^{-\beta\cdot t}$}
 
    \addplot[
    color=red,
    mark=diamond,
    mark options={solid},
    style=dashed
    ] table [x=time, y=simul, col sep=space] {cdf_10nodes.txt};\addlegendentry{\Large Simulation}

    \end{axis}
\end{tikzpicture}\caption{ $|S_l|=10$}
\end{subfigure}
}
\caption{Fitting curve to CDF of the first meeting time obtained from simulation.}
\label{fig:cdfMeetingTime}
\end{figure}
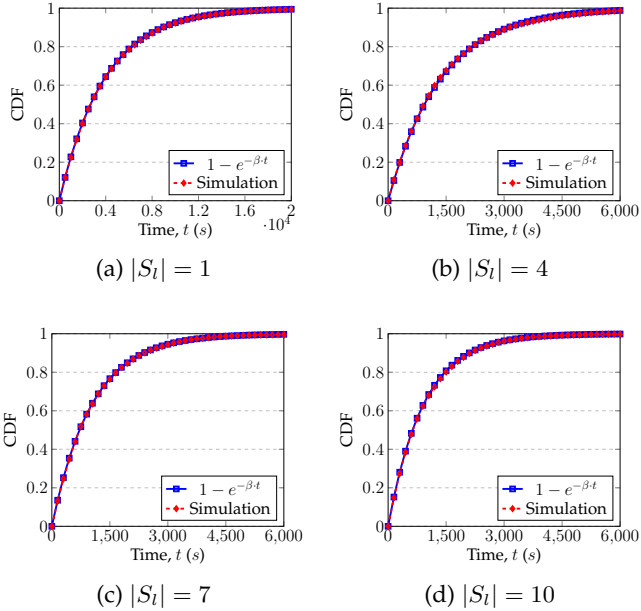

Let $L=1000~m$, $L_c=100~m$, $R=10~m$, $v_{min}=5~m/s$, $v_{max}=15~m/s$, $\alpha^{-1}=80~s$, and $\beta^{-1}=520~s$ as \cite{4215676} and \cite{Hsu:2009:MST:1665838.1665854}. Assume that there is a community, denoted by $c$, centered at $(250,250)$ considering the left-lower corner of the common area as origin. Fig.~\ref{fig:cdfMeetingTime} represents the results obtained from simulation for CDFs of the first meeting time of $A_r$ with a set of nodes, $S_l$, moving in community~$c$ for $|S_l|=1$, $4$, $7$, and $10$. Note that the case of $|S_l|=1$ corresponds to the first meeting time of a roaming node and a local node. In Fig.~\ref{fig:cdfMeetingTime}, each curve is obtained from 10,000 independent runs of simulation.
As it can be seen in Fig.~\ref{fig:cdfMeetingTime}, the curve $1-e^{-\beta\cdot t}$ fits simulation results and the CDF of the first meeting time exhibits exponential behavior. We use a Chi-Square test \cite{greenwood1996guide} to analyze how the CDF of the exponential distribution fits the CDF of the first meeting time of $A_r$ with $S_l$. Considering the number of bins and the significance level 40 and 0.01, respectively, the results of Chi-square test corresponding to Figures~\ref{fig:cdfMeetingTime}(a), (b), (c), and (d) are 15.60, 51.64, 31.63, and 28.26, respectively. All the results of Chi-square test are less than the critical value 62.43 indicating that the simulation results match the CDF of an exponential distribution.

\begin{figure}
\centering
\begin{tikzpicture}[scale=0.45]
\begin{axis}[
    xlabel={\LARGE  $n$},
    ylabel={\LARGE $R_{meet}(n)$~($s^{-1}$)},
    xmin=1, xmax=10,
    ymin=0, ymax=0.0012,
    xtick={1,2,3,4,5,6,7,8,9,10},
    ytick={0,0.0003,0.0006,0.0009,0.0012},
    ymajorgrids=true,
    grid style=dashed,
     every axis plot/.append style={ultra thick},
     every tick label/.append style={font=\Large}
]

    \addplot[
    color=red,
    mark=square,
    mark options={solid},
    style=dashed
    ]
    coordinates {
        (1,0.0002586	)
        (2,	0.000463030)
        (3,0.000613131	)
        (4,0.000735581	)
        (5,0.000826595	)
        (6,0.000910946 )
        (7,0.000962966 )
        (8, 0.001013051)
        (9, 0.001047914)
        (10, 0.001086226)

    };

\end{axis}
\end{tikzpicture}
\caption{The values of $R_{meet}(n)$ for $1\leq n\leq 10$.} 
\label{fig:R_meet}
\end{figure}
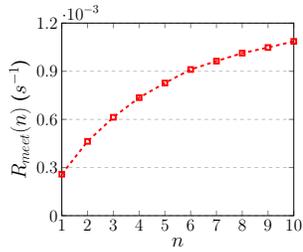

In order to perform further analysis, let $n=|S_l|$ and $R_{meet}(n)$, $n\ge 1$, denote the rate of the exponential distribution representing the first meeting time of node $A_r$ with a node of $S_l$. The events of meetings of node $A_r$ with the nodes belonging to $S_l$ are not independent since all of the nodes belonging to $S_l$ move in the same community. If node $A_r$ meets one of them, the probability of meeting any other node of $S_l$ after a short time increases. Thus, $R_{meet}(n)$ does not equal to $n\cdot R_{meet}(1)$, which is confirmed by simulation results. For example, Fig.~\ref{fig:R_meet} represents function $R_{meet}(n)$ for $n=1,2,\dots, 10$, values were obtained which is obtained from simulation of a network with $L=1000~m$, $L_c=100~m$, $R=10~m$, $v_{min}=5~m/s$, $v_{max}=15~m/s$, $\alpha^{-1}=80~s$, and $\beta^{-1}=520~s$. As observed in Fig.~\ref{fig:R_meet}, $R_{meet}(n)$ does not linearly increase with $n$.

\section{overview of previous models}\label{sec:overview}
This section presents an overview of the monolithic and folded SRN models proposed in our previous work \cite{8843121}. Due to the strict limitation of space, we do not present details of SRNs. The formal definition and structure of SRNs can be found in \cite{peterson1981petri, ajmone1984class, muppala1991composite, ajmone1995modelling, bause2002stochastic}.

The previous monolithic model consists of $N+1$ submodels, one per each community and one to represent the state of the destination node. Submodel $Sub_i$, $1\leq i\leq N$, of the previous monolithic model represents the situation of nodes frequently visiting community $c_i$ excluding the destination node, $i=N$. Submodel $Sub_i$ contains four main places to represent the infected local nodes, the infected roaming nodes, the susceptible local nodes, and the susceptible roaming nodes that frequently visit community~$c_i$. Excluding the initial number of tokens, submodels $Sub_i$, \mbox{$1\leq i\leq N$}, have the same structures.

In order to be able to evaluate the performance of large-scale networks, we have proposed a folded model by folding submodels $Sub_1$, $Sub_2$, \dots, $Sub_{N-1}$ together into a single submodel, named $Sub_f$. In addition to the places representing nodes and the transitions representing infections and decision of nodes about the movement mode, there are another place acting as a counter, named $P_{cnt}$, and two other transitions in submodel $Sub_f$. These elements enumerate the number of communities, among $c_1$, $c_2$, \dots, $c_{N-1}$, which are frequently visited by at least one infected node.

In contrast to the monolithic model, the number of susceptible (infected) nodes frequently visiting each community, except community $c_N$, cannot be captured from the folded model. Moreover, the number of local infected nodes and the number of local susceptible nodes in each community are not represented in the folded model. However, the values of these quantities are needed to precisely define guard and rate functions of some timed transitions of the folded model.
In order to overcome this shortcoming, we use an approximation. If there are $k$ tokens in place~$P_{cnt}$, due to symmetry, we assume that at least one infected node frequently visits communities $c_1$, $c_2$, \dots, $c_k$. The approximation is based on the assumption that nearly the same number of infected nodes and the same number of roaming infected nodes frequently visit each community $c_j$, $1\leq j\leq k$.

\section{The Proposed Monolithic Model}\label{sec:mono}
In this section, we describe the proposed monolithic SRN to evaluate the average and the CDF of delivery delay and the average number of transmissions of the epidemic routing in the network model described in Section~\ref{sec:networkModel}. In addition to $N$, $M$, $\alpha$, $\beta$, $P_r$, $P_{l}$, and $P_{sel\_i}$, $1\leq i\leq N$, introduced in Section~\ref{sec:networkModel}, the proposed monolithic model has the following input parameters. These parameters are the rates of the exponential functions at which first meeting times are distributed.
\begin{itemize}
    \item $\lambda$: The rate of the first meeting time of any two local nodes which move in the same community 
    \item $\mu$: The rate of the first meeting time of any two roaming nodes
    \item $\gamma$: The rate of the first meeting time of any roaming node with any local node ($R_{meet}(1)$)
    \item $\eta$: The rate of the first meeting time of a roaming node with the set of other nodes when they are in local mode and move in the same community ($R_{meet}(M-1)$)
\end{itemize}

\begin{figure}
    \centering
   \resizebox{0.9\columnwidth}{!}{
    \includegraphics[]{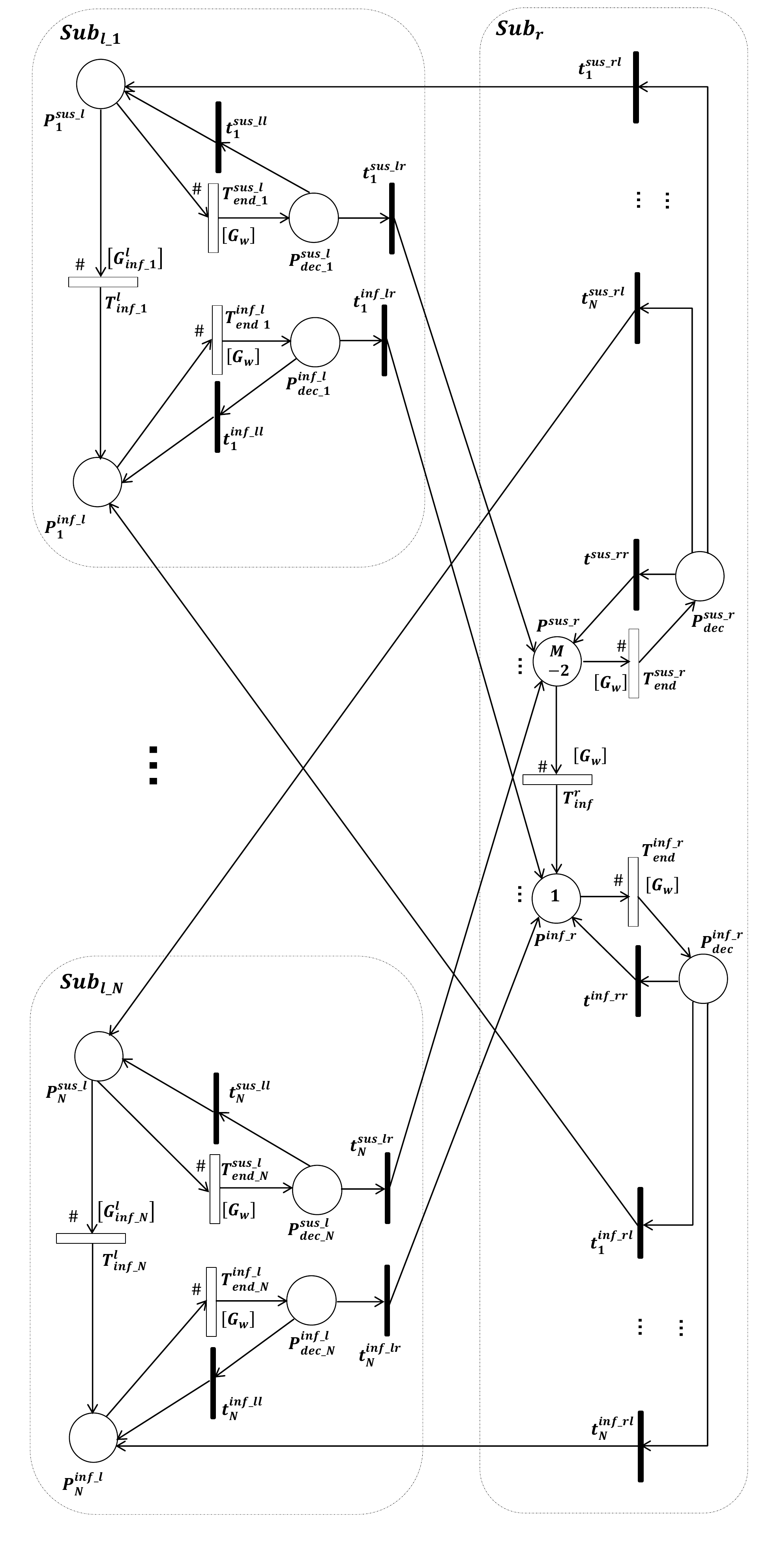}}
    \caption{Submodels $Sub_r$, $Sub_{l\_1}$, \dots, $Sub_{l\_N}$ of the proposed monolithic model.}
    \label{fig:monoModel}
\end{figure}

\begin{figure}
    \centering
   \resizebox{0.9\columnwidth}{!}{
    \includegraphics[]{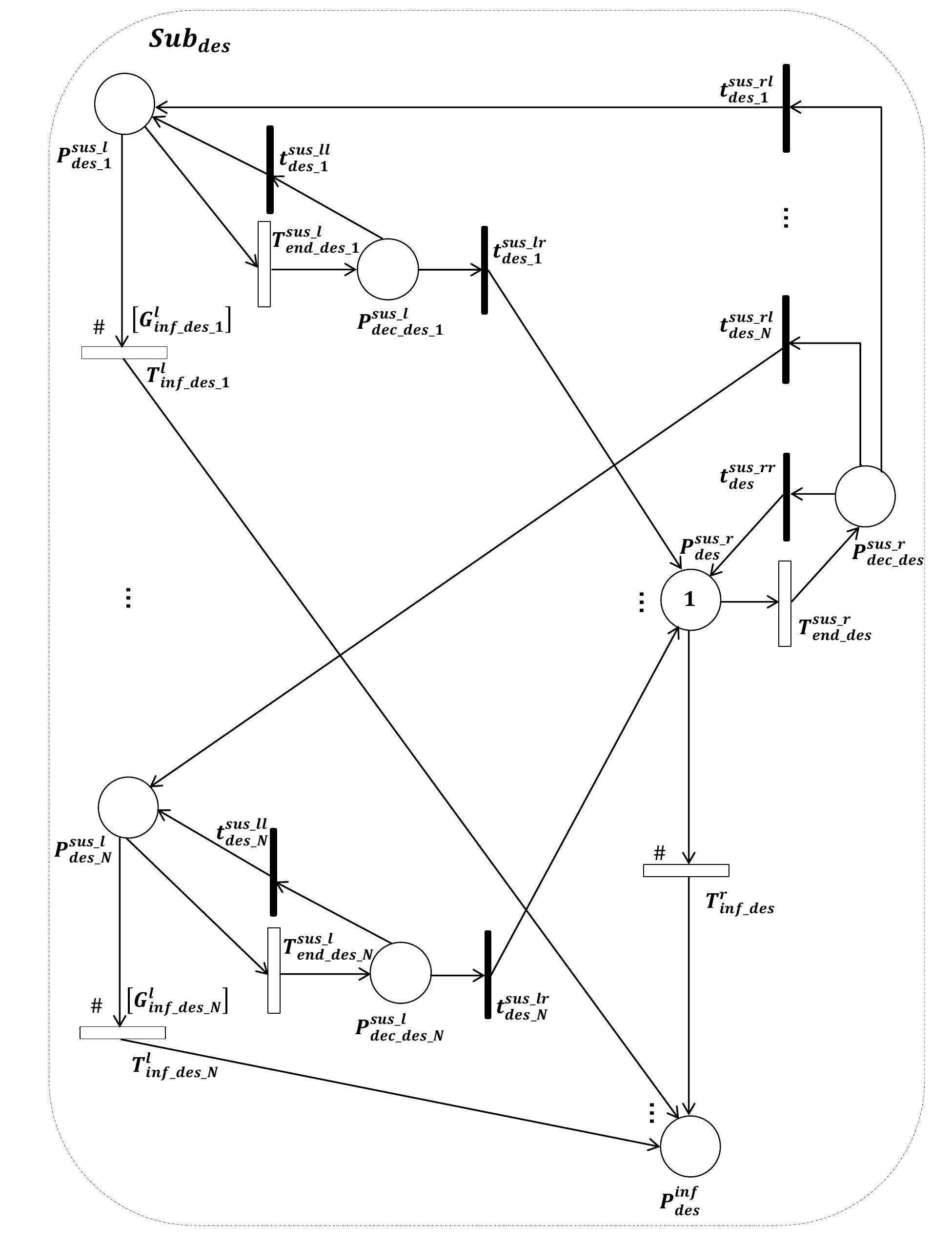}}
    \caption{Submodel $Sub_{des}$ of the proposed monolithic model.}
    \label{fig:dest}
\end{figure}

The proposed monolithic model has $N+2$ submodels, named $Sub_{l\_1}$, $Sub_{l\_2}$, \dots, $Sub_{l\_N}$, $Sub_{r}$, and $Sub_{des}$. Submodels $Sub_{l\_1}$, $Sub_{l\_N}$, and $Sub_{r}$ are represented in Fig.~\ref{fig:monoModel}. Submodels $Sub_{l\_j}$, \mbox{$1<j<N$}, have the same graphical representation as $Sub_{l\_1}$ and $Sub_{l\_N}$. Thus, these submodels are not shown in Fig.~\ref{fig:monoModel}.
As in the previous models, submodel $Sub_{des}$, represented in Fig.~\ref{fig:dest}, captures the situation of the destination node. Excluding the destination node, submodel $Sub_{l\_j}$, $1\leq j\leq N$, represents the nodes that are in local mode and move in community~$c_j$ while submodel $Sub_{r}$ represents roaming nodes. In the following, first, the role of elements of each submodel is described, and then the guard and rate functions of the transitions are introduced. It is worth mentioning that the range of $j$ in the rest of the paper is from 1 to $N$ ($1\leq j\leq N$). 

\subsection{Description of Elements}\label{sec:desOfMonoElements}
Place~$P^{sus\_l}_j$ ($P^{inf\_l}_j$) of submodel $Sub_{l\_j}$ contains the tokens representing the susceptible (infected) local nodes that are moving in community $c_j$. Transition $T^l_{inf\_j}$ represents the infection of a susceptible local node while moving in community $c_j$. Transition $T^{sus\_l}_{end\_j}$ ($T^{inf\_l}_{end\_j}$) represents the ending of travels of susceptible (infected) local nodes moving in community~$c_j$. When transition $T^{sus\_l}_{end\_j}$ ($T^{inf\_l}_{end\_j}$) fires, a token is removed from place $P^{sus\_l}_j$ ($P^{inf\_l}_j$) and put into place $P^{sus\_l}_{dec\_j}$ ($P^{inf\_l}_{dec\_j}$). As soon as a token is put in place $P^{sus\_l}_{dec\_j}$ ($P^{inf\_l}_{dec\_j}$), one of transitions $t^{sus\_ll}_j$ and $t^{sus\_lr}_j$ ($t^{inf\_ll}_j$ and $t^{inf\_lr}_j$) fires with probabilities $1-P_r$ and $P_r$, respectively. Transitions $t^{sus\_ll}_j$ and $t^{sus\_lr}_j$ ($t^{inf\_ll}_j$ and $t^{inf\_lr}_j$) represent choosing local and roaming modes, respectively, by the susceptible (infected) node, that has just finished its travel in local mode. 

Places $P^{sus\_r}$ and $P^{inf\_r}$ are containers for tokens representing the susceptible and infected roaming nodes, respectively. According to the second assumption provided in Section~\ref{sec:networkModel}, and given that initially only the source node has the message, the initial numbers of tokens in places $P^{sus\_r}$ and $P^{inf\_r}$ are $M-2$ and 1, respectively. Transition $T^r_{inf}$ represents the infection of a susceptible roaming node. Moreover, transition $T^{sus\_r}_{end}$ ($T^{inf\_r}_{end}$) models the end of travel of a susceptible (infected) roaming node. As soon as a token is put in places $P^{sus\_r}_{dec}$ and $P^{inf\_r}_{dec}$, it is removed upon firing of an immediate transition. Both transitions $t^{sus\_rr}$ and $t^{inf\_rr}$ fire with the probability $1-P_l$, which represents remaining in the roaming mode during the next travel. If a token is in place $P^{sus\_r}_{dec}$ ($P^{inf\_r}_{dec}$), with probability $P_l$, one of transitions $t^{sus\_rl}_1$, $t^{sus\_rl}_2$, \dots, $t^{sus\_rl}_N$ ($t^{inf\_rl}_1$, $t^{inf\_rl}_2$, \dots, $t^{inf\_rl}_N$) fires. Specifically, transitions $t^{sus\_rl}_j$ and $t^{inf\_rl}_j$ represent that the node which has just finished its travel in roaming mode, selects the local mode and moving in community $c_j$. Thus, these transitions fire with probability $P_l\cdot P_{sel\_j}$.  

As it can be seen in Fig.~\ref{fig:dest}, there is one initial token in place $P^{sus\_r}_{des}$ of submodel $Sub_{des}$. This token represents the destination node and circulates among places of this submodel until it is put in place $P^{inf}_{des}$. When this token is in place $P^{sus\_l}_{des\_j}$ ($P^{sus\_r}_{des}$), the destination node is in community $c_j$ (common area) and moves in local (roaming) mode. Depositing of this token in place $P^{inf}_{des}$ represents the delivery of message to the destination node. Roles of place $P^{sus\_l}_{dec\_des\_j}$ and transitions $T^l_{inf\_des\_j}$, $T^{sus\_l}_{end\_des\_j}$, $t^{sus\_ll}_{des\_j}$, $t^{sus\_lr}_{des\_j}$, $t^{sus\_rl}_{des\_j}$ are similar to those of place $P^{sus\_l}_{dec\_j}$ and transitions $T^l_{inf\_j}$, $T^{sus\_l}_{end\_j}$, $t^{sus\_ll}_{j}$, $t^{sus\_lr}_{j}$, $t^{sus\_rl}_{j}$ of submodel $Sub_{l\_j}$, respectively. Moreover, place $P^{sus\_r}_{dec\_des}$ and transitions $T^{sus\_r}_{end\_des}$, $t^{sus\_rr}_{des}$, and $T^r_{inf\_des}$ can be described in a similar manner to the place $P^{sus\_r}_{dec}$ and transitions $T^{sus\_r}_{end}$, $t^{sus\_rr}$, and $T^r_{inf}$ of submodel $Sub_r$, respectively. The only difference of the aforementioned elements of submodel $Sub_{des}$ with those of submodels $Sub_{l\_j}$ and $Sub_r$ is that the elements of $Sub_{des}$ represent the situation of the destination node exclusively, while corresponding elements of $Sub_{l\_j}$ and $Sub_r$ are used to model the situation of all other susceptible nodes.

\subsection{Guard and rate functions}


As mentioned in Section~\ref{sec:desOfMonoElements}, the existence of a token in place~$P^{inf}_{des}$ indicates that the message is delivered to the destination. Since the average delivery delay is one of our measures of interest, the proposed monolithic model is designed to be absorbed when the token representing the destination is put in place $P^{inf}_{des}$. To this end, a guard function satisfying condition $\#P^{inf}_{des}== 0$ should be associated with each timed transition in submodels $Sub_{l\_j}$ and $Sub_r$.

We associate the guard function $G_w$, defined as Eq.~(\ref{eq:G_w}), to all timed transitions in submodel~$Sub_{l\_j}$ and $Sub_r$ except transitions $T^{l}_{inf\_j}$.
\begin{equation} \label{eq:G_w}
G_w= (\#P^{inf}_{des}== 0)
\end{equation}
In addition to condition $\#P^{inf}_{des}== 0$, there is another condition which should be satisfied before firing the transition~$T^{l}_{inf\_j}$. Each local node has a chance to meet only roaming nodes and other local nodes of the community in which it moves. Thus, we associate the guard function $G^{l}_{inf\_j}$, defined as Eq.~(\ref{eq:G_inf}), to transition~$T^{l}_{inf\_j}$ to guarantee that there is at least one infected local node in community~$c_j$ or at least one infected roaming node in the common area.
\begin{equation} \label{eq:G_inf}
G^{l}_{inf\_j}= \big(\#P^{inf}_{des}== 0~\wedge~(\#P^{inf\_l}_{j}+\#P^{inf\_r})>0\big)
\end{equation}
Guard function $G^{l}_{inf\_j}$ is also associated with transition $T^{l}_{inf\_des\_j}$ in submodel $Sub_{des}$.

In order to precisely compute the rates of transitions $T^l_{inf\_j}$ and $T^{r}_{inf}$, value of function $R_{meet}$ is required.
However, we use the values of this function only for $n=1$ and $n=M$ as input parameters to simplify the proposed model, and approximate function $R_{meet}(n)$ as linear function $\hat{R}_{meet}(n)$, defined as Eq.~(\ref{eq:R_meet}).
\begin{equation}\label{eq:R_meet}
    \hat{R}_{meet}(n)=
    \left\{
	\begin{array}{ll} 0, & n=0\\
	\gamma, & n=1\\
	\gamma+(n-1)\cdot\frac{\eta-\gamma}{M-2}, & n>1\\
		\end{array}
\right.
\end{equation} 

Each susceptible local node in community $c_j$ meets each infected local node in that community with rate $\lambda$. As mentioned earlier, the number of tokens in place $P^{sus\_l}_j$ ($P^{inf\_l}_j$) represents the number of susceptible (infected) local nodes in community $c_j$. The meeting rate of these susceptible and infected nodes is \mbox{$\#P^{sus\_l}_j\times \#P^{inf\_l}_j\times \lambda$}. Moreover, the time taken for each infected roaming node to meet the first susceptible local node that moves in community~$c_j$ is distributed with rate \mbox{$\hat{R}_{meet}(\#P^{sus\_l}_j)$}. The number of infected roaming nodes is given by $\#P^{inf\_r}$. Therefore, the rate of transition $T^l_{inf\_j}$ can be computed by Eq.~(\ref{eq:Rlinf}).
\begin{equation}\label{eq:Rlinf}
    R^l_{inf\_j}= \#P^{sus\_l}_j\cdot \#P^{inf\_l}_j\cdot \lambda + \#P^{inf\_r}\cdot \hat{R}_{meet}(\#P^{sus\_l}_j)
\end{equation}
If the destination moves in a community during local mode, it meets each infected local node moving in that community with rate $\lambda$. Moreover, the destination meets each infected roaming node with rate $\gamma$. Thus, the rate of transition $T^l_{inf\_des\_j}$ is computed by Eq.~(\ref{eq:Rlinfdes}).
\begin{equation}\label{eq:Rlinfdes}
    R^l_{inf\_des\_j}=\#P^{inf\_l}_j\cdot\lambda+\#P^{inf\_r}\cdot\gamma
\end{equation}

Each susceptible roaming node meets each infected roaming node with rate $\mu$. Moreover, the time taken for each susceptible roaming node to meet the first infected local node in community $c_j$ is distributed with rate $\hat{R}_{meet}(\#P^{inf\_l}_j)$. The number of susceptible roaming nodes is equal to $\#P^{sus\_r}$. Therefore, the rate of transition $T^{r}_{inf}$ is computed by Eq.~(\ref{eq:Rrinf}).
\begin{equation}\label{eq:Rrinf}
    R^r_{inf}= \#P^{sus\_r}\cdot\big(\#P^{inf\_r}\cdot \mu + \sum^{N}_{j=1}\hat{R}_{meet}(\#P^{inf\_l}_j)\big)
\end{equation}
In a similar manner to transition~$T^r_{inf}$, the rate of transition~$T^r_{inf\_des}$ is obtained from Eq.~(\ref{eq:Rrinf_des}).
\begin{equation}\label{eq:Rrinf_des}
    R^r_{inf\_des}= \#P^{inf\_r}\cdot \mu + \sum^{N}_{j=1}\hat{R}_{meet}(\#P^{inf\_l}_j)
\end{equation}

The duration of the travel is exponentially distributed with rates $\alpha$ and $\beta$ in local and roaming modes, respectively. Thus, transitions $T^{sus\_l}_{end\_des\_j}$ and $T^{sus\_r}_{end\_des\_j}$ fire with rates $\alpha$ and $\beta$, respectively. The rates of transitions $T^{sus\_l}_{end\_j}$, $T^{inf\_l}_{end\_j}$, $T^{sus\_r}_{end}$ and $T^{inf\_r}_{end}$ are computed by Eqs.~(\ref{eq:Rulend}) to (\ref{eq:Rirend}), respectively.
\begin{equation}\label{eq:Rulend}
    R^{sus\_l}_{end\_j}=\#P^{sus\_l}_j\cdot \alpha
\end{equation}
\begin{equation}\label{eq:Rilend}
    R^{inf\_l}_{end\_j}=\#P^{inf\_l}_j\cdot \alpha
\end{equation}
\begin{equation}\label{eq:Rurend}
    R^{sus\_r}_{end}=\#P^{sus\_r}\cdot \beta
\end{equation}
\begin{equation}\label{eq:Rirend}
    R^{inf\_r}_{end}=\#P^{inf\_r}\cdot \beta
\end{equation}

\section{The Proposed Folded Approximate Model}\label{sec:folded}
\begin{figure}
    \centering
   \resizebox{0.9\columnwidth}{!}{
    \includegraphics[]{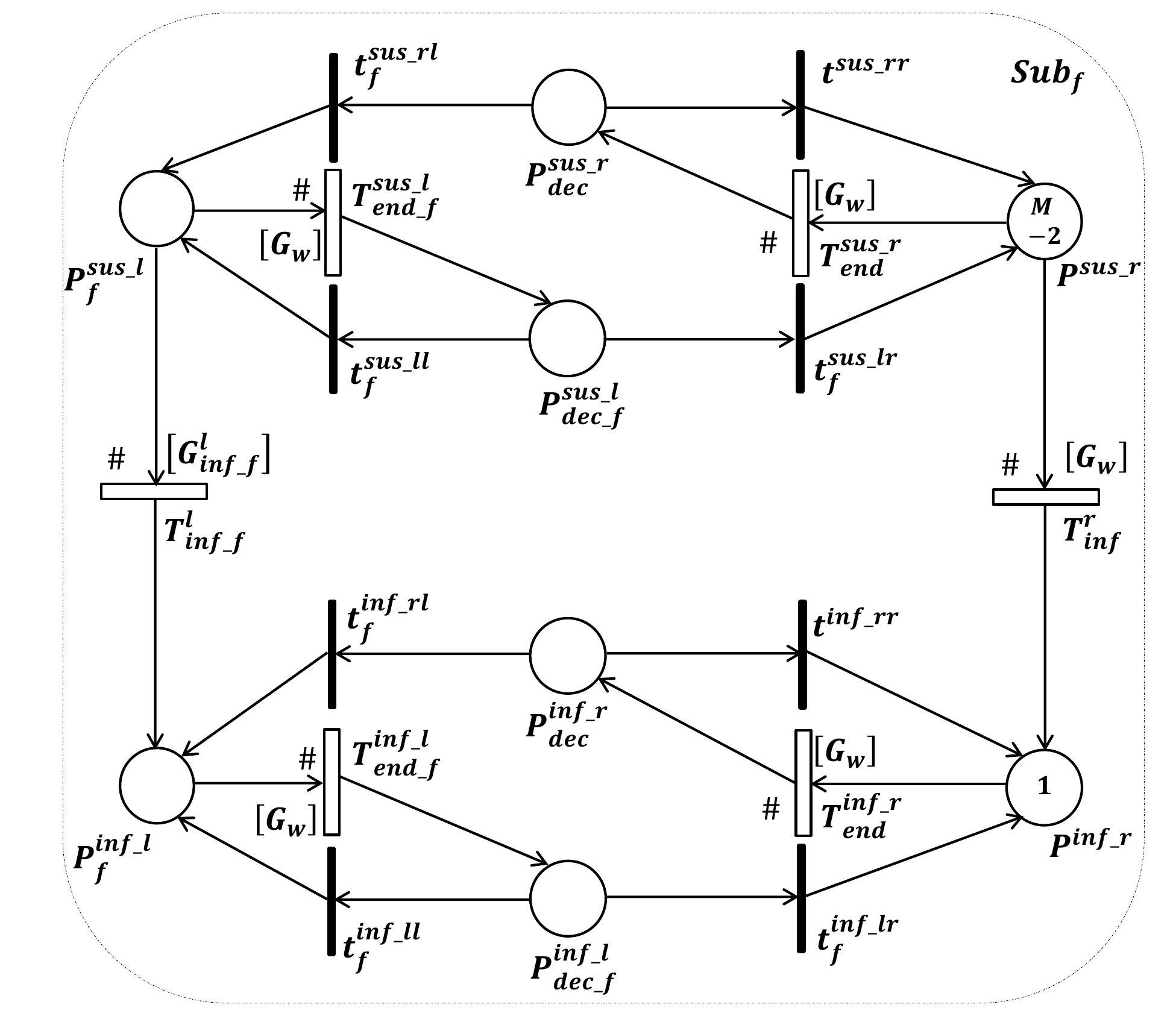}}
    \caption{Submodel $Sub_f$ of the proposed folded model.}
    \label{fig:foldedSubModel}
\end{figure}

Containing at least four places per each community among which tokens representing local nodes circulate, the monolithic model is not scalable in terms of $N$, the number of communities, and $M$, the number of nodes. To overcome this difficulty, in this section, we propose a folded approximate model to evaluate the performance of epidemic routing in the target mobile social networks. In contrast with the monolithic model, in the folded model there is one submodel, named $Sub_f$, instead of submodels $Sub_{l\_j}$ and $Sub_{r}$. As mentioned in Section~\ref{sec:mono}, submodels $Sub_{l\_1}$, $Sub_{l\_2}$, \dots, $Sub_{l\_N}$ of the monolithic model have the same structure. Thus,
in order to prevent rapid growth of the state space, we fold submodels $Sub_{l\_1}$, $Sub_{l\_2}$, \dots, $Sub_{l\_N}$ all together. Since places $P^{sus\_l}_j$ ($P^{inf\_l}_j$) are folded into a single place, we need to fold transitions $t^{sus\_rl}_j$ ($t^{inf\_rl}_j$) of submodel $Sub_r$ into a single transition. Submodel $Sub_f$, represented in Fig.~\ref{fig:foldedSubModel}, results from applying the folding technique on submodels $Sub_{l\_j}$ and some elements of submodel $Sub_r$ of the monolithic model, and then merging the elements resulting from the folding with elements of submodel $Sub_r$ that are not folded. Table~\ref{tab:correspondence} provides details of submodel~$Sub_f$ elements in the folded model. For each element that results from folding, Table~\ref{tab:correspondence} shows the corresponding elements of the monolithic model that are folded. The
initial number of tokens of each place, rate functions of timed transitions and firing probabilities of immediate transitions are included in Table~\ref{tab:correspondence}.



\begin{table}
	\renewcommand{\arraystretch}{1.4}
	\setlength{\tabcolsep}{1pt}
	\begin{center}
	\caption{Elements of the proposed folded submodel}
	\label{tab:correspondence}
	\end{center}
	
	\centering
	\begin{tabular}{c|c|c}
		\hline
		{\bfseries Element of} & {\bfseries Corresponding} & {\bfseries Initial Mark /} \\ 
		\hhline{~~~}
		{\bfseries Submodel} $\boldsymbol{Sub_f}$ & {\bfseries Folded Elements in}& {\bfseries Rate Function /}\\
		\hhline{~~~}
		& {\bfseries the monolithic model} &  {\bfseries Firing Probability}\\
		\hline
		\hline
		 
		$P^{sus\_l}_f$  & $P^{sus\_l}_j$ ($1\leq j\leq N$) & 0\\
		\hline
		  $P^{sus\_r}$ & - & $M-2$\\
		\hline
        $P^{inf\_l}_f$ & $P^{inf\_l}_j$ ($1\leq j\leq N$) & 0\\
        
        \hline
        $P^{inf\_r}$ &   - &  1\\ 
        \hline
         $P^{sus\_r}_{dec}$ & -  & 0\\
	     \hline
	     $P^{sus\_l}_{dec\_f}$ & $P^{sus\_l}_{dec\_j}$ ($1\leq j\leq N$) & 0\\
	     \hline
	    $P^{inf\_r}_{dec}$ &  - & 0\\
	     \hline
	       $P^{inf\_l}_{dec\_f}$ &  $P^{inf\_l}_{dec\_j}$ ($1\leq j\leq N$) & 0\\
	      \hline
	       $T^{sus\_l}_{end\_f}$ & $T^{sus\_l}_{end\_j}$ ($1\leq j\leq N$) & $R^{sus\_l}_{end\_f}$\\
	      \hline
	       $t^{sus\_rl}_{f}$ & $t^{sus\_rl}_{j}$ ($1\leq j\leq N$) & $P_l$\\
	      \hline
	       $t^{sus\_ll}_{f}$ & $t^{sus\_ll}_{j}$ ($1\leq j\leq N$) & $1-P_r$\\
	      \hline
        $T^{sus\_r}_{end}$ &  - & $R^{sus\_r}_{end}$ \\ 
	      \hline
	       $t^{sus\_rr}$ & - & $1-P_l$ \\
	      \hline
	      $t^{sus\_lr}_{f}$ &  $t^{sus\_lr}_{j}$ ($1\leq j\leq N$) & $P_r$\\
	      \hline
	      $T^l_{inf\_f}$ &  $T^l_{inf\_j}$ ($1\leq j\leq N$) & $R^l_{inf\_f}$\\
	      \hline
	      $T^r_{inf}$ & - & $R^r_{inf}$\\
	      \hline
	      
	       $T^{inf\_l}_{end\_f}$ &  $T^{inf\_l}_{end\_j}$ ($1\leq j\leq N$) & $R^{inf\_l}_{end\_f}$\\
	      \hline
	       $t^{inf\_rl}_{f}$ & $t^{inf\_rl}_{j}$ ($1\leq j\leq N$) & $P_l$\\
	      \hline
	       $t^{inf\_ll}_{f}$ & $t^{inf\_ll}_{j}$ ($1\leq j\leq N$) & $1-P_r$\\
	      \hline
         $T^{inf\_r}_{end}$ & - & $R^{inf\_r}_{end}$\\ 
	      \hline
	       $t^{inf\_rr}$ & - & $1-P_l$ \\
	      \hline
	      $t^{inf\_lr}_{f}$ & $t^{inf\_lr}_{j}$ ($1\leq j\leq N$) &  $P_r$\\
	      \hline
	\end{tabular}
\end{table}

\begin{figure}
    \centering
   \resizebox{0.9\columnwidth}{!}{
    \includegraphics[]{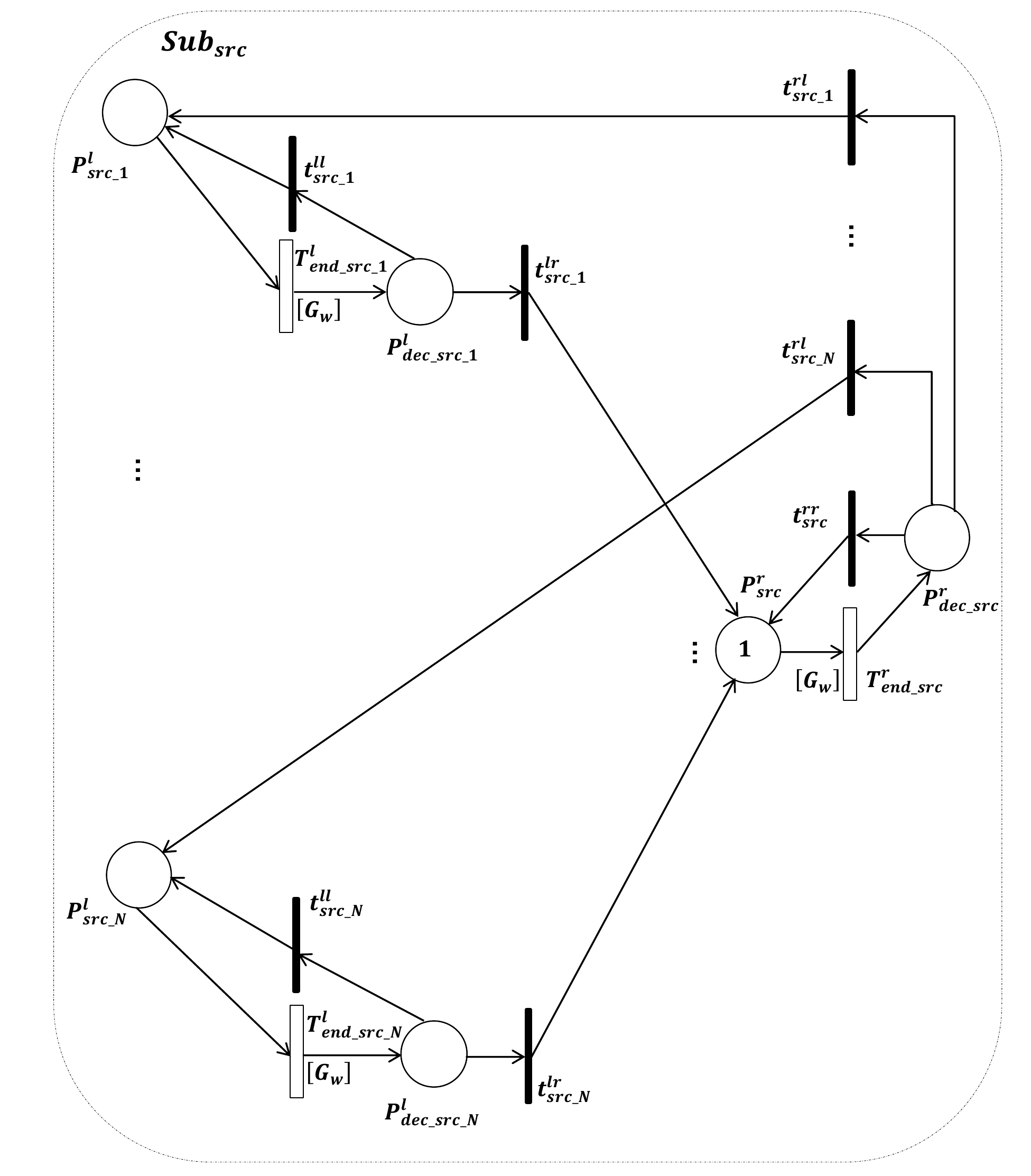}}
    \caption{Submodel $Sub_{src}$ of the proposed folded model.}
    \label{fig:src}
\end{figure}

In submodel $Sub_f$, only places $P^{sus\_l}_f$ and $P^{inf\_l}_f$ are used to represent the susceptible and infected local nodes, respectively. Thus, this submodel does not capture the number of local infected (susceptible) nodes moving in a specific community.
Since the source is the only initial infected node in the network, capturing the community in which it moves during local mode by the analytical model, is important to achieve a good accuracy when the probabilities of selecting communities are not equal, and there are a few nodes in the network. Under these conditions, the community in which the source node moves during local mode has a significant effect on the average time at which the first infection occurs. Thus, we model the situation of the source node in a specific submodel, named $Sub_{src}$, which is represented in Fig.~\ref{fig:src}. 

There exists an initial token in place $P^r_{src}$ which represents the source node and circulates among places of submodel $Sub_{src}$. The existence of the token in place $P^{l}_{src\_j}$ ($P^{r}_{src}$) represents that the source node is in community $c_j$ (common area) and moves in local (roaming) mode. Roles of place $P^{l}_{dec\_src\_j}$ and transitions $T^{l}_{end\_src\_j}$, $t^{ll}_{src\_j}$, $t^{lr}_{src\_j}$, $t^{rl}_{src\_j}$ are similar to those of place $P^{inf\_l}_{dec\_j}$ and transitions $T^{inf\_l}_{end\_j}$, $t^{inf\_ll}_{j}$, $t^{inf\_lr}_{j}$, $t^{inf\_rl}_{j}$ of submodel $Sub_{l\_j}$ of the monolithic model, respectively. Moreover, place $P^{r}_{dec\_src}$ and transitions $T^{r}_{end\_src}$ and $t^{rr}_{src}$ can be described in a similar manner with place $P^{inf\_r}_{dec}$ and transitions $T^{inf\_r}_{end}$ and $t^{inf\_rr}$ of submodel $Sub_r$ of the monolithic model, respectively. The only difference between the aforementioned elements of submodel $Sub_{src}$ and those of submodels $Sub_{l\_j}$ and $Sub_r$ is that elements of $Sub_{src}$ represent the situation of the source node exclusively, but corresponding elements of $Sub_{l\_j}$ and $Sub_r$ model the situation of all infected nodes except the destination node. It is worth mentioning that transitions $t^{ll}_{src\_j}$, $t^{lr}_{src\_j}$, $t^{rl}_{src\_j}$, and $t^{rr}_{src}$
fire with probabilities $1-P_r$, $P_r$, $P_l\cdot P_{sel\_j}$, and $1-P_l$, respectively.

In addition to submodels $Sub_f$ and $Sub_{src}$, the proposed folded model has another submodel, named $Sub_{des}$, to represent the situation of the destination node as the monolithic model. Submodel $Sub_{des}$ of the folded model has the same graphical representation as the submodel $Sub_{des}$ of the monolithic model that is represented in Fig.~\ref{fig:dest}. Thus, the elements of $Sub_{des}$ are not described herein. 

\subsection{Guard and Rate Functions}
Transitions $T^{sus\_l}_{end\_f}$, $T^{sus\_r}_{end}$, $T^{inf\_l}_{end\_f}$, and $T^{inf\_r}_{end}$ represent the end of travels of the susceptible local nodes, susceptible roaming nodes, infected local nodes, and infected roaming nodes, respectively, excluding the source and destination nodes. 
Since the duration of each travel of local (roaming) nodes is exponentially distributed with rate~$\alpha$ ($\beta$), the rates of these transitions are computed by Eqs.~(\ref{eq:R_sus_l_end_f}) to (\ref{eq:R_inf_r_end_f}).
\begin{equation}\label{eq:R_sus_l_end_f}
    R^{sus\_l}_{end\_f}=\#P^{sus\_l}_{f}\cdot\alpha
\end{equation}
\begin{equation}
     R^{sus\_r}_{end}=\#P^{sus\_r}\cdot\beta
\end{equation}
\begin{equation}
     R^{inf\_l}_{end\_f}=\#P^{inf\_l}_{f}\cdot\alpha
\end{equation}
\begin{equation}\label{eq:R_inf_r_end_f}
     R^{inf\_r}_{end}=\#P^{inf\_r}\cdot\beta
\end{equation} Moreover, rates of transition $T^l_{end\_src\_j}$ of submodel $Sub_{src}$ and transition $T^{sus\_l}_{end\_des\_j}$ of submodel $Sub_{des}$ are $\alpha$ while transition $T^r_{end\_src}$ of submodel $Sub_{src}$ and transition $T^{sus\_r}_{end\_des}$ of submodel $Sub_{des}$ fire with rate $\beta$.
Similarly to the corresponding transitions in the monolithic model, the guard function $G_w$, defined in Eq.~(\ref{eq:G_w}), is associated with all timed transitions of submodels~$Sub_{src}$ and $Sub_f$, except transition~$T^l_{inf\_f}$ to construct an absorbing model.

Places $P^{sus\_l}_f$ and $P^{inf\_l}_f$ act as repositories for tokens representing susceptible and infected local nodes, respectively.
Thus,
in contrast with the monolithic model, the number of susceptible and infected local nodes in each community cannot be captured from the folded model. However, the values of these quantities are needed to precisely define guard and rate functions of the timed transitions of submodels $Sub_f$ and $Sub_{des}$ that represent infection of nodes. To overcome this difficulty, we apply the approximation in Algorithm~\ref{alg:approx}. Let $\hat{N}^{inf\_l}_i$ ($\hat{N}^{sus\_l}_i$), $1\leq i\leq N$, denote an approximated number of infected (susceptible) local nodes moving in community $c_i$. Algorithm~\ref{alg:approx} can be used to compute both approximated values $\hat{N}^{inf\_l}_i$ and $\hat{N}^{sus\_l}_i$ based on the number of tokens in places $P^{inf\_l}_f$, $P^{sus\_l}_f$, $P^{l}_{src\_i}$, and $P_{sel\_i}$ ($1\leq i\leq N$). Since the procedures to compute $\hat{N}^{inf\_l}_i$ and $\hat{N}^{sus\_l}_i$ are similar, Algorithm~\ref{alg:approx} is written in a generic form where $x$ can be \textit{inf} or \textit{sus}.
\begin{algorithm}[t]
\small
\setstretch{0.9}
\KwData{$\#P^{x\_l}_f$, $P_{sel\_i}$, $\#P^{l}_{src\_i}$ ($1\leq i\leq N$)}
\KwResult{$\hat{N}^{x\_l}_i$~($1\leq i\leq N$)} 
\nl
$\hat{N}^{x\_l}_i=   \nint{ P_{sel\_i} \cdot \#P^{x\_l}_f}$ ($1\leq i\leq N$);\\
\nl
$d=\sum_{i=1}^{N} \hat{N}^{x\_ l}_{i} - \#P^{x\_l}_f $;\\
\nl
\For{$i=1 \ to \  N$}{
\nl \If{$ P_{sel\_i}\cdot \#P^{x\_l}_f  > \hat{N}^{x\_l}_i$}{
  \nl add $c_i$ to $Q^{+}$.
   }\nl \Else{
\nl   add $c_i$ to $Q^{-}$.
  }
  	}
\nl
Sort\ $Q^{+}$ and $Q^{-}$\ based\ on\ measure $\lvert P_{sel\_i}\cdot\#P^{x\_l}_f  - \hat{N}^{x\_l}_i \rvert$\ in\  descending\ order.\\
\nl \While{$d> 0$}{
\nl		$c_k=$~remove front element of $Q^{-}$;\\
\nl		$\hat{N}^{x\_l}_{k} --$;\\
\nl		$d--$;\\
   }
\nl \While{$d< 0$}{
\nl		$c_k=$~\text{remove front element of} $Q^{+}$;\\
\nl		$\hat{N}^{x\_l}_{k} ++$;\\
\nl		$d++$;\\
    }
\nl \If{$x==inf$}{
\nl \For{$i=1 \ to \  N$}{
\nl \If{$\#P^{l}_{src\_i}==1$}{
\nl     $\hat{N}^{x\_l}_{i}++$; \\
\nl		$break$;
		}
		}
    }
\caption{Approximation of the number of infected local nodes and the number of susceptible local nodes in each community $c_i$ (\mbox{$1\leq i\leq N$}), excluding the destination node}
 \label{alg:approx}
\end{algorithm}

In each marking of the folded SRN, there are $\#P^{inf\_l}_f$ and $\#P^{sus\_l}_f$ tokens that represent the infected and susceptible local nodes, respectively.
Algorithm 1 distributes the $\#P^{inf\_l}_f$ ($\#P^{sus\_l}_f$) tokens representing the infected (susceptible) local nodes among communities $c_1$, \dots, $c_{N}$ based on the probabilities $P_{sel\_ 1}$, \dots, $P_{sel\_ N}$, according to which prospective local nodes select communities. $P_{sel\_i}\times\#P^{inf\_l}_f$ ($P_{sel\_i}\times \#P^{sus\_l}_f$) can be a good indicator of the approximate number of infected (susceptible) local nodes in community $c_i$. However, this indicator may not be an integer. If that is the case, we round the indicator to the nearest integer, and then we assign as many nodes as that integer to each community. In Algorithm 1, $\nint{P_{sel\_i}\cdot \#P^{x\_l}_f}$ represents the integer nearest to $P_{sel\_i}\cdot \#P^{x\_l}_f$. We define $d$ as follows,
\begin{equation}
    d=\sum_{i=1}^N   \nint{ P_{sel\_i}\cdot\#P^{x\_l}_f} - \#P^{x\_ l}_f.
\end{equation}
$d$ may not be zero. If that is the case, we need to revise node assignments. In the following, we show how to revise the node assignments for each cases of $d>0$ and $d<0$.
\begin{enumerate}
    \item $\boldsymbol{d>0}$\textbf{:} We deallocate $d$ nodes from the communities, for which their indicators were rounded-up, such that at most one node from each of these communities can be deallocated. We call this part of the algorithm \textit{deallocation phase}.
    
    \item $\boldsymbol{d<0}$\textbf{:} We assign $-d$ more nodes to communities, for which their indicators were rounded-down, such that at most one more node to each of these communities can be allocated. We call this part of the algorithm \textit{reallocation phase}. 
\end{enumerate}

Deallocation and reallocation phases are performed using two priority queues of communities, denoted by $Q^{-}$ and $Q^{+}$, respectively. $Q^{-}$ ($Q^{+}$) is a queue of communities, for which their indicators were rounded-up (rounded down), sorted in descending order based on measure \mbox{$|( P_{sel\_i}\cdot \#P^{x\_l}_f ) -   \nint{P_{sel\_i} \cdot\#P^{x\_l}_f}|$} for each community $c_i$. During deallocation (reallocation) phase, we start from the head of $Q^{-}$ ($Q^{+}$) and deallocate (allocate) one node from (to) each community until $|d|$ nodes are deallocated (allocated). At the end, if the source node is in the local mode and moves in community $c_i$, we increase $\hat{N}^{inf\_l}_i$ by one.

In addition to guard function $G_w$, we need to define other guard functions to be associated to transitions~$T^l_{inf\_f}$ and $T^l_{inf\_des\_j}$. Guard function $G^l_{inf\_f}$, defined by Eq.~(\ref{eq:G_l_inf}), is associated with transition $T^l_{inf\_f}$. Transition~$T^l_{inf\_f}$ models meetings of the local susceptible nodes, excluding the destination node, with infected nodes. Such a meeting is possible only if at least one infected node moves in the roaming mode, or movement modes of one susceptible node, except the destination node, and one infected node are local, and they move in the same community. This condition is guaranteed by $G^l_{inf\_f}$ as represented in Eq.~(\ref{eq:G_l_inf}). Furthermore, this guard function guarantees that the destination has not received the message as $G_w$.
\begin{equation}\label{eq:G_l_inf}
\begin{split}
    &G^l_{inf\_f}= (\#P^{inf}_{des}==0)~\wedge\\
    &\big( (\sum_{j=1}^{N}\hat{N}^{sus\_l}_j\cdot \hat{N}^{inf\_l}_j>0)~\vee~(\#P^{inf\_r}+\#P^{r}_{src}>0) \big)
\end{split}
\end{equation}
In a similar way, we associate guard function $G^l_{inf\_des\_j}$, defined by Eq.~(\ref{eq:G_l_inf_des_j}), to transition $T^l_{inf\_des\_j}$.
\begin{equation}\label{eq:G_l_inf_des_j}
    G^{l}_{inf\_des\_j}= (\hat{N}^{inf\_l}_j>0)~\vee~(\#P^{inf\_r}+\#P^{r}_{src}>0)
\end{equation} Note that roaming susceptible nodes have chance to meet each infected node, and there is at least one infected node, the source, always in the network. Thus, function $G_w$ is an appropriate guard function for transition $T^{r}_{inf}$, and we do not need to include any condition regarding the numbers of infected and susceptible nodes.

Each susceptible local node in community $c_j$ meets each infected local node moving in that community with rate~$\lambda$. The time taken for each infected roaming node to meet one of the susceptible local nodes in community~$c_j$ is distributed with rate $\hat{R}_{meet}(\hat{N}_j^{sus\_l})$. The tokens in places $P^{inf\_r}$ and $P^{r}_{src}$ are the only tokens that represent the infected roaming nodes. Thus,
the rate of transition $T^l_{inf\_f}$ is computed as Eq.~(\ref{eq:R_l_inf}).
\begin{equation}\label{eq:R_l_inf}
\begin{split}
    R^l_{inf\_f}&=\sum_{j=1}^{N}\Big(\hat{N}^{sus\_l}_j\cdot \hat{N}^{inf\_l}_j\cdot \lambda\\
    &+(\#P^{inf\_r}+\#P^{r}_{src})\cdot \hat{R}_{meet}(\hat{N}^{sus\_l}_j)\Big)
\end{split}
\end{equation} If movement mode of the destination node is local, it meets each infected roaming node with rate $\gamma$ and each infected local node in the community, in which the destination node moves with rate $\lambda$. Thus, the rate of transition $T^{l}_{inf\_des\_j}$ is obtained from Eq.~(\ref{eq:R_l_inf_des_j}).
\begin{equation}\label{eq:R_l_inf_des_j}
    R^l_{inf\_des\_j}=(\#P^{inf\_r}+\#P^r_{src})\cdot\gamma+\hat{N}^{inf\_l}_j\cdot\lambda
\end{equation}

Every susceptible roaming node meets every infected roaming node with rate $\mu$. Moreover, the rate of the time taken for a susceptible roaming node to meet one of infected local nodes, in a community, is estimated using function $\hat{R}_{meet}$. Thus, the rates of transitions $T^{r}_{inf\_des}$ and $T^r_{inf}$ are computed by Eqs.~(\ref{eq:R_r_inf_des}) and (\ref{eq:R_r_inf}), respectively.
\begin{equation}\label{eq:R_r_inf_des}
    R^r_{inf\_des}=(\#P^{inf\_r}+\#P^{r}_{src})\cdot \mu + \sum_{i=1}^N \hat{R}_{meet}(\hat{N}^{inf\_l}_i)
\end{equation}
\begin{equation}\label{eq:R_r_inf}
    R^r_{inf}=\#P^{sus\_r}\cdot\Big( (\#P^{inf\_r}+\#P^{r}_{src})\cdot \mu + \sum_{i=1}^N \hat{R}_{meet}(\hat{N}^{inf\_l}_i)\Big)
\end{equation}

\section{Measures of interest}\label{sec:measures}
In this section, the performance measures and the way to compute them, by applying the proposed models, are presented.

\textit{Average Delivery Delay:} As mentioned in Sections \ref{sec:mono} and \ref{sec:folded}, the proposed monolithic and folded models are absorbed when a token is put in place $P^{inf}_{des}$, representing the delivery of the message to the destination. Thus, the \textit{Mean Time To Absorption} (\textit{MTTA}) in both monolithic and folded models represents the average delivery delay.

\textit{Average Number of Transmissions:} This measure can be computed from the proposed models after an appropriate reward rate is assigned to each tangible marking of SRNs. According to Section \ref{sec:desOfMonoElements}, tokens representing the infected nodes, except the destination, circulate among places $P^{inf\_l}_j$ and $P^{inf\_r}$ of the monolithic model. Thus, in each marking of the monolithic SRN, the sum of the numbers of tokens in these places represents the number of infected nodes, excluding the destination while including the source, that is equal to the number of transmissions. If we assign the reward rate represented in Eq.~(\ref{eq:reward_mono}) to each marking of the monolithic SRN, the average reward rate at time $t$ is equal to the average number of transmissions until time $t$.
\begin{equation}\label{eq:reward_mono}
    r_{m}=\sum_{j=1}^N \#P^{inf\_l}_j+\#P^{inf\_r}
\end{equation}
As $t$ increases, the average reward rate at time $t$ converges to the average number of transmissions by the delivery time. Thus, if $t$ is large enough, the average number of transmissions by time of delivery is obtained.
Similarly, the average number of transmissions can be obtained from the folded model. Tokens representing the infected nodes excluding the source and destination nodes are hold in places $P^{inf\_l}_f$ and $P^{inf\_r}$. Thus,
the appropriate reward rate, to be assigned to markings of the folded SRN, is obtained from Eq.~(\ref{eq:reward_folded}). Note that addition of 1 in this equation is due to counting also the transmission of the message to the destination node.
\begin{equation}\label{eq:reward_folded}
    r_{f}=\#P^{inf\_l}_f+\#P^{inf\_r}+1
\end{equation}

\textit{CDF of the Delivery Delay:} The probability of delivery of the message no later than time $t$, $t>0$, can be computed using transient analysis of the proposed SRNs. To this end, we need to assign the reward rate $\#P^{inf}_{des}$ to each marking of the SRNs. The CDF of delivery delay at time $t$ is equal to the average reward rate at time $t$.


\section{Performance evaluation}\label{sec:perf}
This section presents the results obtained from both the monolithic and folded models, and these results are validated. Moreover, we propose an ODE model for epidemic routing in the target network, and then compare it with the folded model in terms of accuracy.
We set the network parameters as in \cite{4215676} and \cite{Hsu:2009:MST:1665838.1665854}. Specifically, parameters $L$, $L_c$, $P_l$, and $P_r$ are set to 1000 $m$, 100 $m$, $0.8$, and $0.2$, respectively. Moreover, the average duration of a travel in local and roaming modes is 80~$s$ and 520~$s$, ($\alpha=1/80$ and $\beta=1/520$), respectively. According to \cite{4215676} and \cite{Hsu:2009:MST:1665838.1665854}, these setting matches the MIT trace \cite{Balazinska:2003:CMN:1066116.1066127}. It is worth noting that parameters $R$, $v_{min}$, $v_{max}$, and $v_{trans}$ are 10~$m$, 5~$m/s$, 15~$m/s$, and 20~$m/s$, respectively. Table~\ref{tab:communities} represents the locations of communities and the probabilities at which they are selected by the prospective local nodes for different values of $N$. Considering the left-lower corner of common area as the origin of a coordinate system, communities are centered at the coordinates mentioned in Table~\ref{tab:communities}.
\begin{table}
\small
	\renewcommand{\arraystretch}{1}
	\setlength{\tabcolsep}{0.5pt}
	\begin{center}
	\caption{Locations of communities and the probabilities of choosing them by prospective local nodes.}
	\label{tab:communities}
	\end{center}
	\centering
	\begin{tabular}{c|c|c}
		\hline
		\textbf{Community ($c_i$)} & 
		 \textbf{Coordinate of Center} & $\boldsymbol{P_{sel\_i}}$\\
		
		\hline
		 \hline
		 \multicolumn{3}{c}{$\boldsymbol{N=3}$}\\
		 \hline
		$c_1$ & (250,~250) & 0.2\\
		  \hline
		   	$c_2$ & (250,~750) & 0.4\\
		   	\hline
		   	$c_3$ & (750,~250) & 0.4\\
		   	\hline
		   	 \multicolumn{3}{c}{$\boldsymbol{N=4}$}\\
		 \hline
		 	$c_1$ & (250,~250) & 0.2\\
		  \hline
		   	$c_2$ & (250,~750) & 0.4\\
		   	\hline
		   	$c_3$ & (750,~250) & 0.1\\
		   	\hline
		   	$c_4$ & (750,~750) & 0.3\\
		   	\hline
		   	 \multicolumn{3}{c}{$\boldsymbol{N=5}$}\\
		 \hline
		 	$c_1$ & (250,~250) &  0.2\\
		  \hline
		   	$c_2$ & (250,~750) & 0.4\\
		   	\hline
		   	$c_3$ & (750,~250) & 0.2\\
		   	\hline
		   	$c_4$ & (750,~750) & 0.1\\
		   	\hline
		   	$c_5$ & (500,~500) & 0.1\\
		   	\hline
	\end{tabular}
\end{table}

Before using the proposed models to evaluate the performance of epidemic routing, we need to compute the input parameters, namely $\lambda$, $\mu$, $\gamma$, and $\eta$, by simulation. In order to obtain $\lambda$ ($\mu$), in each run of the simulation, two nodes are uniformly placed in an $L_c\times L_c$ ($L\times L$) square, and then nodes are moved until they meet each other. Similarly, in order to compute $\gamma$, one node is uniformly placed in an $L\times L$ square, and one node is placed uniformly in an $L_c\times L_c$ square within the aforementioned $L\times L$ square. Then, the former and latter nodes move in the $L\times L$ and $L_c\times L_c$ squares, respectively, until they meet each other. Parameter $\eta$ is obtained in a similar way when computing~$\gamma$. After computing parameters $\lambda$, $\mu$, $\gamma$, and $\eta$, the proposed models were numerically solved, and performance metrics computed by using the SPNP \cite{ciardo1989spnp}. This tool automatically converts an SRN to its underlying CTMC and facilitates computing the measures of interest. 

\begin{figure}
\begin{tikzpicture}
\begin{axis}[
    xlabel={Time,~\textit{t}~(\textit{s})},
    ylabel={CDF of the Delivery Delay},
    xmin=1, xmax=1600,
    ymin=0, ymax=1,
    xtick={0,300,600,900,1200,1500},
    ytick={0,0.1,0.2,0.3,0.4,0.5,0.6,0.7,0.8,0.9,1},
    ymajorgrids=true,
    legend pos=north west,
    grid style=dashed,
     every axis plot/.append style={ultra thick}
]
 

    \addplot[
    color=green,
    mark=diamond,
    mark options={solid}
    ]
    coordinates {
        (0,0)
        (100, 0.034090933)
        (200, 0.089312219)
        (300, 0.164876696 )
        (400, 0.256099631	)
        (500, 0.356499802 )
        (600, 0.458988854 )
        (700,0.557178433 )
        (800, 0.646324082 )
        (900, 0.723664062 )
        (1000, 0.78824447)
        (1100, 0.84046644)
        (1200, 0.881572581 )
        (1300,0.913205474)
        (1400, 0.937089742)
        (1500,0.954836584)
        (1600,0.967845427)
        (1700, 0.977271962)

    };\addlegendentry{Monolithic Model}

    \addplot[
    color=blue,
    mark=triangle,
    mark options={solid},
    style=dashed
    ]
    coordinates {
         (0,0)
        (100,0.032125 )
        (200,0.089625 )
        (300, 0.16975)
        (400, 0.273125	)
        (500, 0.3745 )
        (600,0.484875 )
        (700, 0.59475)
        (800, 0.67125)
        (900, 0.74175 )
        (1000,0.811375)
        (1100,0.860125 )
        (1200, 0.898125)
        (1300,0.928)
        (1400,0.947375)
        (1500,0.966875)
        (1600,0.978875)
        (1700, 0.98925)

    };\addlegendentry{Simulation}

\end{axis}
\end{tikzpicture}
\caption{The CDF of the delivery delay obtained from the proposed monolithic SRN model and the simulation when \mbox{$N=4$} and \mbox{$M=15$}.} 
\label{fig:CDF}
\end{figure}
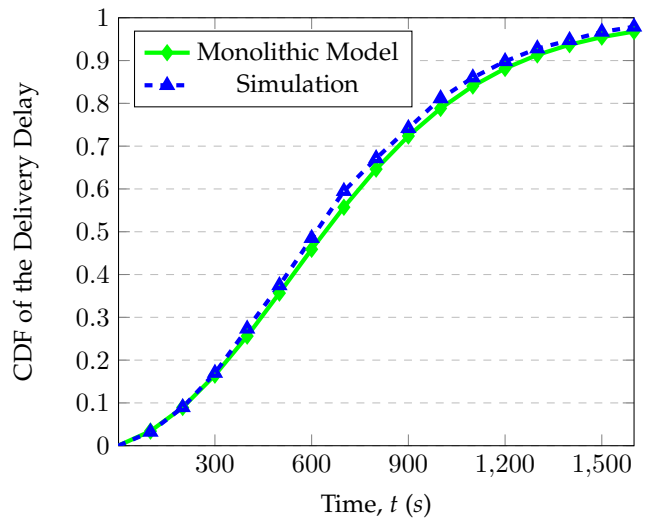

In order to validate the proposed models, we compare the results of the monolithic and folded models against the simulation results. To achieve this end, the network under-analysis is simulated by applying the discrete-event simulation developed in Java. Although the transmission delays are not considered in the proposed models, they are considered in the simulation. We assume that the transmission speed of each node and the message size are 2.5~\textit{MBps} and 25~\textit{KB}, respectively. This transmission speed could be provided by Bluetooth technology. Prior to presenting the numerical results, it is worth mentioning that each simulation result, reported in the rest of this section, is calculated as the average of the values obtained from 8000 independent runs. Fig.~\ref{fig:CDF} represents the CDF of the delivery delay obtained from the monolithic model and the simulation for the aforementioned network with four communities ($N=4$) and 15 nodes ($M=15$). As can be seen in Fig.~\ref{fig:CDF}, simulation and analytical results are very close to each other indicating high accuracy of the monolithic model to predict CDF of the delivery delay.

Table~\ref{tab:mono_fold} represents the average delivery delay and the average number of transmissions obtained from the proposed monolithic model and the simulation. Columns \textit{Mono.} and \textit{Sim.} represent the results of the monolithic model and the simulation, respectively. Moreover, the Percent Errors of the results computed from the monolithic model with respect to the corresponding results computed from the simulation are represented in columns~\textit{PE}. Due to the memory shortage, the monolithic model cannot be solved for some configurations. Notation \mbox{"-"} in Table~\ref{tab:mono_fold} shows these configurations, where the monolithic model encounters a scalability problem in terms of the number of communities, $N$, and the number of nodes, $M$. As an example, for networks consisting of 4 communities ($N=4$), a system with 64 \textit{GB} memory space cannot solve the monolithic model even when there are only 20 nodes in the network ($M=20$). As it can be seen in Table~\ref{tab:mono_fold}, the results obtained from the monolithic model and the simulation are close to each other indicating that the monolithic SRN accurately models the network. Moreover, the average number of transmissions does not depend on the number of communities, $N$, and consequently the location visiting preference. According to the results represented in Table~\ref{tab:mono_fold}, the average number of transmissions is nearly $M/2$. This is in accordance with our prior work \cite{8843121} where the average number of transmissions can be estimated as a linear function of the total number of nodes. 

\begin{table}
	\renewcommand{\arraystretch}{0.9}
	\setlength{\tabcolsep}{4pt}
	\begin{center}
	\caption{Comparison of the results obtained with the monolithic model and by simulation: Mono - monolithic model; Sim - simulation; PE - Percent Errors.}
	\label{tab:mono_fold}
	\end{center}
	\centering
	\begin{tabular}{c|c|c|c|c|c|c|c}
		\hline
		\multirow{3}{*}{$\boldsymbol{N}$} & 
		\multirow{3}{*}{$\boldsymbol{M}$} &  \multicolumn{3}{c|}{\bfseries Average Delivery} & \multicolumn{3}{c}{\bfseries Average Number of}\\
		\hhline{~~~~~~~~}
		& &  \multicolumn{3}{c|}{\bfseries Delay ($s$)} & \multicolumn{3}{c}{\bfseries Transmissions}\\
		\hhline{~~------}
		& & \bfseries  {Mono.}  & \bfseries {Sim.}  & \bfseries  {PE} & \bfseries  {Mono.}  & \bfseries {Sim.}  & \bfseries  {PE}\\
		\hline
		\hline
		{\bfseries 3} & {\bfseries 5} & 1272.72 & 1253.50 & 1.53 & 2.50 & 2.51 & 0.39\\
		\hhline{~-------}
		& {\bfseries 10} &  845.69 &  821.60 & 2.93 & 4.95 & 4.96 & 0.14 \\
		\hhline{~-------}
		& {\bfseries 15} & 671.34 &  639.88 & 4.91 & 7.39 & 7.47 & 1.12\\
		\hhline{~-------}
	    & {\bfseries 20} & 570.16 & 536.05 & 6.36 & 9.82 & 9.97 & 1.52\\
	    \hhline{~-------}
		\hline
		{\bfseries 4} & {\bfseries 5} & 1366.03 & 1346.21 & 1.47 & 2.50 & 2.50 & 0.23 \\
		\hhline{~-------}
	     & {\bfseries 10} & 892.31 & 874.31 & 2.06 & 4.96 & 5.02 & 1.34 \\
        \hhline{~-------}
	     \bfseries & {\bfseries 15} &  700.22 & 668.48 & 4.75 & 7.39 & 7.48 & 1.78\\
	    \hhline{~-------}
	     \bfseries & {\bfseries 20} & - & 552.78 & - & - & 10.06 & -\\
	    \hline
	     {\bfseries 5} & {\bfseries 5} & 1442.32 & 1416.52 & 1.82 & 2.50 & 2.50 & 0.03 \\
	    \hhline{~-------}
	     & {\bfseries 10} & 931.34 &  906.67 & 2.72 & 4.96 & 5.02 & 1.30\\
        \hhline{~-------}
	     \bfseries &  {\bfseries 15} & 724.65 & 696.87 & 3.99 & 7.39 & 7.52 & 1.69\\
	    \hhline{~-------}
	     \bfseries &  {\bfseries 20} & - & 573.95 & - & - & 10.09 & -\\
	     \hline

	\end{tabular}
\end{table}

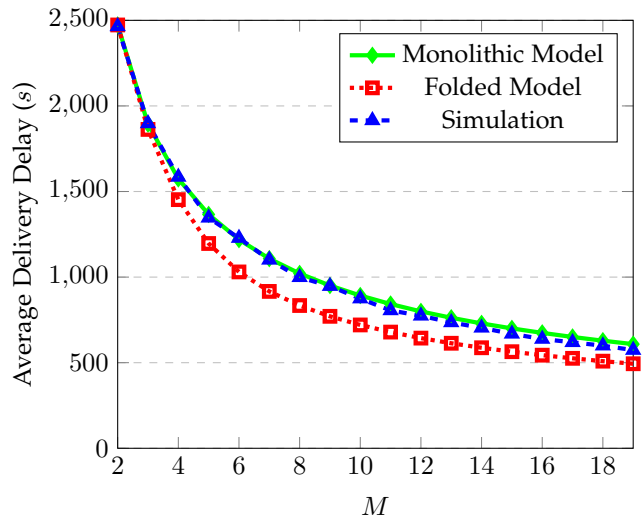
\begin{figure}
\centering
\begin{tikzpicture}
\begin{axis}[
    xlabel={$M$},
    ylabel={Average Delivery Delay~($s$)},
    xmin=2, xmax=19,
    ymin=0, ymax=2500,
    xtick={2,4,6,8,10,12,14,16,18},
    ytick={0,500,1000,1500,2000,2500},
    ymajorgrids=true,
    legend pos=north east,
    grid style=dashed,
     every axis plot/.append style={ultra thick},
]
 

    \addplot[
    color=green,
    mark=diamond,
    mark options={solid}
    ]
    coordinates {
        (2,2472.93179)
        (3,1894.082576)
        (4,1573.777939)
        (5,1366.028229)
        (6,1218.933497)
        (7,1107.808714)
        (8,1021.157089)
        (9,950.6648665)
        (10,892.3051472)
        (11,842.5709379)
        (12,799.5081508)
        (13,762.4024473)
        (14,729.374958)
        (15,700.222482)
        (16,674.1312847)
        (17,650.267119)
        (18, 628.4174539)
        (19,608.3854676)

    };\addlegendentry{Monolithic Model}
    \addplot[
    color=red,
    mark=square,
    mark options={solid},
    style=dotted
    ]    
        coordinates {
        (2,2472.93179)
        (3,1862.772015)
        (4,1452.596626)
        (5,1196.280595)
        (6,1030.189635)
        (7,916.4097258)
        (8,834.0294342)
        (9,770.8825147)
        (10,720.4828887)
        (11,678.8217113)
        (12,643.5776799)
        (13,613.4969323)
        (14,587.2279032)
        (15,564.2315707)
        (16,543.8581588)
        (17, 525.5159992)
        (18, 508.9197893)
        (19, 493.835924)

    };\addlegendentry{Folded Model}
    


    \addplot[
    color=blue,
    mark=triangle,
    mark options={solid},
    style=dashed
    ]
    coordinates {
        (2,2464.614798)
        (3,1897.590474)
        (4,1584.993105)
        (5,1346.212288)
        (6,1227.874099)
        (7,1100.671431)
        (8,998.9129075)
        (9,947.1168137)
        (10,874.3104837)
        (11,805.2329537)
        (12,773.7745862)
        (13,735.1705425)
        (14,703.6220037)
        (15,668.4806837)
        (16,638.2519162)
        (17,619.0545262)
        (18, 599.3004287)
        (19,573.6584075)

    };\addlegendentry{Simulation}

\end{axis}
\end{tikzpicture}
\caption{The average delivery delay obtained from the proposed monolithic and folded SRN
models and the simulation when $N = 4$.}
\label{fig:4com_mono_sim}
\end{figure}

Fig.~\ref{fig:4com_mono_sim} represents the average delivery delay obtained from the monolithic and folded models and by simulation for different values of $M$, when the network under-analysis consists of four communities ($N=4$). Since the monolithic model cannot be solved for \mbox{$M>19$} due to the state space explosion, the maximum value of $M$ is 19 in Fig.~\ref{fig:4com_mono_sim}. As it can be seen in this figure, the results of the monolithic model are very close to the results of the simulation indicating high accuracy of the monolithic model. Moreover, the results of the folded model is close to the results of the simulation and the monolithic model. It indicates that the folded model accurately approximates the monolithic model.  

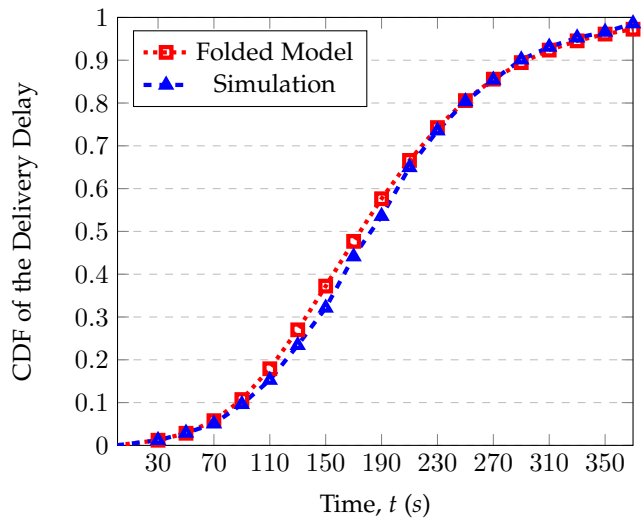
\begin{figure}
\begin{tikzpicture}
\begin{axis}[
    xlabel={Time,~\textit{t}~(\textit{s})},
    ylabel={CDF of the Delivery Delay},
    xmin=1, xmax=370,
    ymin=0, ymax=1,
    xtick={0,30,70,110,150,190,230,270,310,350},
    ytick={0,0.1,0.2,0.3,0.4,0.5,0.6,0.7,0.8,0.9,1},
    ymajorgrids=true,
    legend pos=north west,
    grid style=dashed,
     every axis plot/.append style={ultra thick}
]
 

    \addplot[
    color=red,
    mark=square,
    mark options={solid},
    style=dotted
    ]
    coordinates {
       
        (0,0)
        (30, 0.012037006)
        (50, 0.028303848)
        (70, 0.057771404)
        (90, 0.107110237	)
        (110, 0.179070873 )
        (130, 0.270048842)
        (150,  0.372172034)
        (170, 0.476728936 )
        (190, 0.576361141 )
        (210, 0.665932168 )
        (230, 0.74263856 )
        (250, 0.805694861 )
        (270, 0.855786654 )
        (290, 0.894471255 )
        (310, 0.923667115 )
        (330, 0.945297699)
        (350, 0.961088799)
        (370, 0.972483266)

    };\addlegendentry{Folded Model}

    \addplot[
    color=blue,
    mark=triangle,
    mark options={solid},
    style=dashed
    ]
    coordinates {
        (0,0)
        (30, 0.012)
        (50, 0.029 )
        (70, 0.050625)
        (90, 0.096	)
        (110, 0.152625)
        (130, 0.234 )
        (150, 0.32125 )
        (170, 0.441125)
        (190, 0.535625 )
        (210, 0.649625)
        (230, 0.735375 )
        (250, 0.8035)
        (270, 0.853875)
        (290, 0.90125)
        (310, 0.931)
        (330,0.952625)
        (350, 0.966)
        (370, 0.985625)

    };\addlegendentry{Simulation}

\end{axis}
\end{tikzpicture}
\caption{The CDF of the delivery delay obtained with the proposed Folded SRN model and by simulation when \mbox{$N=4$} and \mbox{$M=100$}.} 
\label{fig:CDF_largeScale}
\end{figure}

In order to evaluate the performance of epidemic routing on a large scale network, the folded model can be adopted. Fig.~\ref{fig:CDF_largeScale} represents the CDF of the delivery delay obtained with the folded model and by simulation for a network with four communities ($N=4$) and 100 nodes ($M=100$). As it can be seen in Fig.~\ref{fig:CDF_largeScale}, simulation and analytical results are very close to each other, indicating high accuracy of the folded model to predict CDF of the delivery delay.

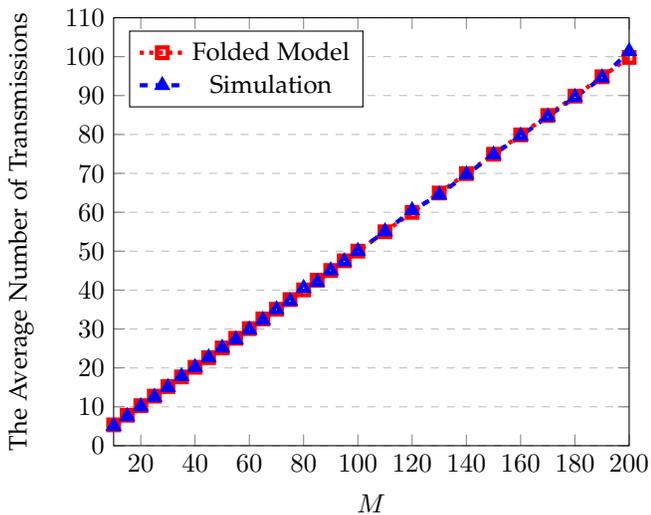
\begin{figure}
\centering
\begin{tikzpicture}
\begin{axis}[
    xlabel={$M$},
    ylabel={The Average Number of Transmissions},
    xmin=10, xmax=200,
    ymin=0, ymax=110,
    xtick={20,40,60,80,100,120,140,160,180,200},
    ytick={0,10,20,30,40,50,60,70,80,90,100,110},
    ymajorgrids=true,
    legend pos=north west,
    grid style=dashed,
     every axis plot/.append style={ultra thick}
]
 

    \addplot[
    color=red,
    mark=square,
    mark options={solid},
    style=dotted
    ]
    coordinates {
        (10,    5.3550969)
        (15,    7.832765773)
        (20,	10.28559892)
        (25,    12.74068595)
        (30,	 15.20464555)
        (35,    17.67714728)
        (40,	20.1560183)
        (45,    22.63864847)
        (50,    25.12341493	)
        (55,    27.60973113)
        (60,    30.0973056)
        (65,    32.5861116742)
        (70,    35.0747451359)
        (75,    37.5640896998)
        (80,    40.0538539681)
        (85,    42.5441074632)
        (90,    45.0335412301)
        (95,    47.5241082771)
        (100,   50.0147847937)
        (110, 54.9956829659)
    (120,59.9766216386)
    (130,64.9573890324)
    (140,69.9382924067)
    (150, 74.9182631579)
    (160,79.89)
    (170,84.876434298)
    (180,89.854498617)
    (190,94.8320745695)
    (200,99.808614743)

    };\addlegendentry{Folded Model}

     \addplot[
    color=blue,
    mark=triangle,
    mark options={solid},
    style=dashed
    ]
    coordinates {
        
    (10,	5.023375)
    (15,	7.52775)    
    (20,	10.061625)
    (25,	12.516375)
    (30,	15.01125)
    (35,	17.81225)   
    (40,	20.150375)
    (45,	22.660375)
    (50,	25.139875)
    (55,	27.262125)
    (60,	29.814)
    (65,	32.28375)
    (70,	35.0315)
    (75,	37.177625)
    (80,	40.494875)
    (85,	42.0415)
    (90,	45.034375)
    (95,	47.339875)
    (100,	49.997875)
    (110, 55.0575)
    (120,60.4905)
    (130,64.488375)
    (140,69.685)
    (150,74.80375)
    (160,79.630625)
    (170,84.590875)
    (180,89.5715)
    (190,94.593)
    (200,101.380875)

    };\addlegendentry{Simulation}

\end{axis}
\end{tikzpicture}
\caption{The average number of transmissions obtained from the proposed folded SRN model and the simulation when \mbox{$N=4$}.} 
\label{fig:4com_folded_transmisson}
\end{figure}

Fig.~\ref{fig:4com_folded_transmisson} represents the average number of transmissions of the message obtained from the folded model and the simulation when the network under-analysis consists of four communities ($N=4$), and $M$ varies from 10 to 200. As it can be seen in Fig.~\ref{fig:4com_folded_transmisson}, the folded model is very accurate to be used for predicting the average number of transmissions. The percent errors corresponding to the results represented in Fig.~\ref{fig:4com_folded_transmisson} is less than 2\% for $M>20$. As shown in Fig.~\ref{fig:4com_folded_transmisson}, the average number of transmissions changes approximately linearly as $M$ increases. Similarly to Table~\ref{tab:mono_fold}, the results obtained from the analytical model and the simulation represented in Fig.~\ref{fig:4com_folded_transmisson} indicates that $M/2$ is an accurate estimation for the average number of transmissions. In order to justify this observation, we present the following theorem about the average number of transmission in a general network model.

\newtheorem{theorem}{Theorem}
\begin{theorem} \label{theorem:1}
The number of transmissions by the time of delivery, including the forwarding to the destination node, follows uniform distribution, in any arbitrary network, where at any time $t$, $t\ge 0$, positions of nodes are independent and have the same PDF.
\end{theorem}
\begin{IEEEproof}
Label the transmissions of the message up to the delivery of the message to the destination node, $T_1$, $T_2$, \dots, $T_{num}$, where $num$ denotes the number of transmissions, and $T_i$ does not occur after $T_j$ \textit{iff} $i<j$. Due to the i.i.d. positions of nodes, the probability of forwarding the message to an arbitrary susceptible node during transmission $T_i$, $1\leq i\leq num$, is $1/(M-i)$. Initially, there exist $M-1$ susceptible nodes in the network. Thus, the destination node receives the message during transmission $T_1$ with the probability $1/(M-1)$.

If only transmissions $T_1$, $T_2$, \dots, $T_{i-1}$, $2\leq i\leq  num$, have occurred, $M-i$ nodes are still susceptible. Thus, if the message has not been forwarded to the destination node in one of the transmissions $T_1$, $T_2$, \dots, $T_{i-1}$, the destination node receives (does not receive) the message during transmission~$T_i$ with probability $1/(M-i)$ ($(M-i-1)/(M-i)$). As a result, the probability of forwarding the message to the destination node during transmission $T_i$ is obtained by Eq.~(\ref{eq:uniformDist}).
\begin{equation}\label{eq:uniformDist}
\begin{split}
    p(i)&= \frac{M-2}{M-1}\times \frac{M-3}{M-2}\times
\cdots \times \frac{M-(i-1)-1}{M-(i-1)}\times \frac{1}{M-i}\\
&=\frac{1}{M-1}.
\end{split}
\end{equation} Note that this proof is valid even if some transmissions occur simultaneously.
\end{IEEEproof}

\newtheorem{corollary}{Corollary}
\begin{corollary} \label{corollary:1}
The average number of transmission by the delivery time in any arbitrary network with $M$ nodes, where at any time $t$, $t\ge 0$, positions of nodes are independent and have the same PDF, is $M/2$.
\end{corollary}
\begin{IEEEproof} According to Theorem~\ref{theorem:1}, the probability of forwarding the message $i$ times, $1\leq i\leq M-1$, by time of delivery including the forwarding to the destination node, is $1/(M-1)$. Thus, the average number of transmissions is computed as,
\begin{equation}
\begin{split}
    \overline{num}&=\frac{1}{M-1}\cdot (1+2+\cdots+ (M-1))\\
    &= \frac{1}{M-1}\cdot \frac{(M-1)\cdot M}{2}=\frac{M}{2}.
\end{split}
\end{equation}
\end{IEEEproof}

Given that initially, nodes are randomly placed within the
common area with a uniform distribution, and they select the communities according to the same PDF, as mentioned in Section~\ref{sec:networkModel}, the herein target network satisfies the condition given in Theorem~\ref{theorem:1}. That is not the case of the network considered in our previous work \cite{8843121} where the community each node frequently visits is different from the communities some other nodes frequently visit. However, the results of Fig.~8 in \cite{8843121} indicate that the average number of transmissions can be estimated as a linear function of the total number of nodes. As the number of transmissions is an important performance measure, it is worth to characterize that when tendencies of nodes to visit a community differ. However, it is challenging, and we leave it for future work.


To the best of our knowledge, there is no analytical approach in the literature, considering exactly the same network model, so we cannot entirely compare the proposed models with the previous approaches. Modeling as ODEs is the main approach to evaluate the performance of DTNs and it was extensively used in the literature \cite{zhang2007performance, banerjee2008relays, ip2008performance, spyropoulos2009routing, hernandez2017analytical}. Hence, we propose an ODE model for epidemic routing in the defined target network, and then compare the accuracy of the proposed folded model with the proposed ODE model. 

We use functions $I^r(t)$, $I^l_i(t)$, and $S^l_i(t)$, $1\leq i\leq N$, to model epidemic routing with ODEs. Let $I^l_i(t)$ and $S^l_i(t)$ denote the average number of infected and susceptible local nodes in community~$c_i$ at time~$t$, respectively, and $I^r(t)$ denote the average number of infected roaming nodes at time~$t$. Thus, the average number of susceptible roaming nodes at time~$t$ can be computed as $M-I^r(t)-\sum_{i=1}^N(I^l_i(t)+S^l_i(t))$. Note that function $\hat{R}_{meet}(n)$, defined in Eq.~(\ref{eq:R_meet}), is a multi-criteria function over $\mathbb{I}$ even when $M>2$. We can define $\hat{R}_{meet}$ as follows,
\begin{equation}\label{R_meet_2}
    \hat{R}_{meet}(n)= \gamma +\theta( 1- n) \cdot (n-1)\cdot \gamma + \theta (n-1)\cdot (n-1)\cdot\frac{\eta-\gamma}{M-2}.
\end{equation} Using functions $I^l_i(t)$, $S^l_i(t)$, and $I^r(t)$ and Eq.~(\ref{R_meet_2}), we model the network described in Section~\ref{sec:networkModel} by $2N+1$ ODEs represented by equations (\ref{eq:I_l}) to (\ref{eq:I_r}) where $1\leq i\leq N$.


\begin{equation}\label{eq:I_l}
\begin{split}
    &\frac{dI^l_i(t)}{dt} = - I^l_i(t)\cdot\alpha\cdot P_r + I^r(t)\cdot\beta\cdot P_l\cdot P_{sel\_i} \\ 
    &+S^l_i(t)\cdot I^l_i(t)\cdot\lambda
    + I^r(t)\cdot \big( \gamma + \theta( 1- S^l_i(t)) \cdot (S^l_i(t)-1)\cdot \gamma\\
    &+ \theta (S^l_i(t)-1)\cdot (S^l_i(t)-1)\cdot\frac{\eta-\gamma}{M-2}
    \big),
\end{split}
\end{equation}
\begin{equation}\label{eq:S_l}
\begin{split}
     &\frac{dS^l_i(t)}{dt}= \big(M-I^r(t)-\sum_{j=1}^N( S^l_j(t)+I^l_j(t))\big)\cdot \beta\cdot P_l\cdot P_{sel\_i}\\
     &-S^l_i(t)\cdot\alpha\cdot P_r-S^l_i(t)\cdot I^l_i(t)\cdot \lambda - I^r(t)\cdot \big( \gamma +\theta( 1- S^l_i(t)) \\
     & \cdot (S^l_i(t)-1)\cdot \gamma + \theta (S^l_i(t)-1)\cdot (S^l_i(t)-1)\cdot\frac{\eta-\gamma}{M-2}\big),
\end{split}
\end{equation}
\begin{equation}\label{eq:I_r}
\begin{split}
     &\frac{dI^r(t)}{dt}=-I^r(t)\cdot\beta\cdot P_l + \sum_{i=1}^N I^l_i(t)\cdot \alpha\cdot P_r+
     \big(M-I^r(t)\\
     &-\sum_{i=1}^N(S^l_i(t)+I^l_i(t))\big)\cdot\Big(I^r(t)\cdot\mu+\sum_{i=1}^{N}\big(\gamma +\theta( 1- S^l_i(t))\cdot  \\
     & (S^l_i(t)-1)\cdot \gamma + \theta (S^l_i(t)-1)\cdot (S^l_i(t)-1)\cdot\frac{\eta-\gamma}{M-2}\big) \Big),
\end{split}
\end{equation} where $\theta$ is the unit step function.

In the network described in Section~\ref{sec:networkModel}, $I^{l}_i(0)$ and $S^{l}_i(0)$, $1\leq i\leq N$, are 0, whereas $I^{r}(0)$ is 1.
Once the system of equations represented in Eqs.~(\ref{eq:I_l}) to (\ref{eq:I_r}) is numerically solved with the aforementioned initial conditions, the average delivery delay, $\mathbb{E}(D)$, is computed by Eq.~(\ref{eq:averageDelay}), as follows \cite{spyropoulos2009routing}.
\begin{equation}\label{eq:averageDelay}
   \mathbb{E}(D)=t_{max}-\frac{\int_{0}^{t_{max}}(I^r(t)+\sum_{i=1}^N I^l_i(t)-1)dt}{M-1},
\end{equation} where $t_{max}$ is a large time such that $I^r(t_{max})+\sum_{i=1}^N I^l_i(t_{max})$ is close to $M$.

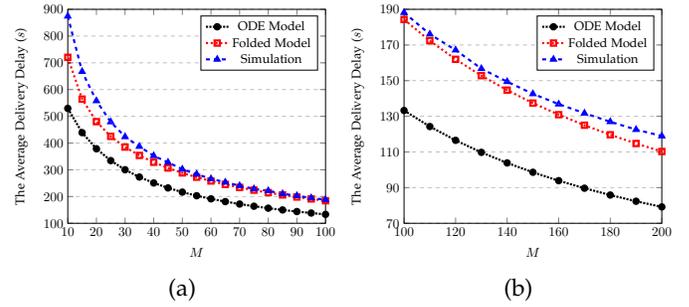
\begin{figure}
\centering
\mbox{
\begin{subfigure}[b]{0.25\textwidth}
\begin{tikzpicture}[scale=0.5]
\begin{axis}[
    xlabel={$M$}, 
    ylabel={The Average Delivery Delay~($s$)},
    xmin=10, xmax=100,
    ymin=100, ymax=900,
    xtick={10,20,30,40,50,60,70,80,90,100},
    ytick={100,200,300,400,500,600,700,800,900},
    ymajorgrids=true,
    legend pos=north east,
    grid style=dashed,
     every axis plot/.append style={ultra thick}
]
 

       \addplot[
    color=black,
    mark=*,
    mark options={solid},
    style=densely dotted
    ]
    coordinates {
    (10, 529.5100	)
    (15,	438.9521)
    (20,  378.7123	)
    (25,  334.3278	 )
    (30,  300.1586	)
    (35,  272.9173	 )
    (40,	  250.7902  )
    (45,  232.2126	   )
    (50,  216.5552	   )
    (55, 203.0191	  )
    (60,   191.3366	 )
    (65,  181.0634	 )
    (70,  171.9474	)
    (75,  163.8104	 )
    (80,  156.4824)
    (85,  149.8476)
    (90,  143.8342)
    (95, 138.3163)
    (100,   133.2473 )

    };\addlegendentry{ODE Model}

    \addplot[
    color=red,
    mark=square,
    mark options={solid},
    style=dotted
    ]
    coordinates {
    (10,	720.4828887)
    (15,	564.2315707)
    (20,	479.991082)
    (25,	424.9484588)
    (30,	384.9420557)
    (35,	353.7161992)
    (40,	328.1565085)
    (45,	306.6892601)
    (50,	288.3032146)
    (55,	272.2643505)
    (60,	258.080785)
    (65,	245.3906129)
    (70,	234.0837286)
    (75,	223.8337608)
    (80,	214.5078117)    
    (85,	205.969036078)
    (90,	198.202355)
    (95,	190.9778534)
    (100,	184.2965947)

    };\addlegendentry{Folded Model}

     \addplot[
    color=blue,
    mark=triangle,
    mark options={solid},
    style=dashed
    ]
    coordinates {
        
    (10,    874.3104837)
    (15,	667.1158125)
    (20,	557.7940775)
    (25,	478.86245)
    (30,	423.3889825)
    (35,	387.2334425)
    (40,	352.4265312)
    (45,	327.7565487)
    (50,	302.0052137)
    (55,	283.5573475)
    (60,	266.3498375)
    (65,	253.3316162)
    (70,	240.4613012)
    (75,	228.4395925)
    (80,	221.5253825)
    (85,	210.3241587)
    (90,	202.916325)
    (95,	194.729725)
    (100, 188.1757025)

    };\addlegendentry{Simulation}

\end{axis}
\end{tikzpicture}
\caption{} 
\end{subfigure}
\hspace{-0.8em}
\begin{subfigure}[b]{0.25\textwidth}
\begin{tikzpicture}[scale=0.5]
\begin{axis}[
    xlabel={$M$}, 
    ylabel={The Average Delivery Delay~($s$)},
    xmin=100, xmax=200,
    ymin=70, ymax=190,
    xtick={100,120, 140, 160, 180, 200},
    ytick={70,90,110,130,150,170,190},
    ymajorgrids=true,
    legend pos=north east,
    grid style=dashed,
     every axis plot/.append style={ultra thick}
]
 

   \addplot[
    color=black,
    mark=*,
    mark options={solid},
    style=densely dotted
    ]
    coordinates {
   
    (100, 	 133.2473) 
    (110,   124.2592) 
    (120,   116.5230)
    (130,   109.7839)
    (140,  103.8601)
    (150,   98.6043) 
    (160,   93.9037)
    (170,   89.6692) 
    (180,  85.8404) 
    (190,  82.3584) 
    (200,  79.1795)

    };\addlegendentry{ODE Model}

    \addplot[
    color=red,
    mark=square,
    mark options={solid},
    style=dotted
    ]
    coordinates {
   
    (100,	184.2965947)
   (110,	 172.356678696)
    (120,	161.953974561)
    (130,	152.802697592)
    (140,	144.672497454)
    (150, 137.421284125)    
    (160,130.900701089)
    (170,125.003711767)
    (180,119.643446247)
    (190,114.745362727)
    (200,110.257270818)

    };\addlegendentry{Folded Model}

     \addplot[
    color=blue,
    mark=triangle,
    mark options={solid},
    style=dashed
    ]
    coordinates {
        
    (100,  188.1757025 )
    (110,	176.0688225)
    (120,	167.1459962)
    (130,	156.7165087)
    (140,	149.412105)
    (150, 142.5881112)
    (160,136.8264537)
    (170,131.7296088)
    (180,126.9408763)
    (190,122.53752)
    (200, 118.9478638)

    };\addlegendentry{Simulation}

\end{axis}
\end{tikzpicture}
\caption{}
\end{subfigure}
}
\caption{The average delivery delay obtained from the proposed folded SRN, ODE model, and the simulation when \mbox{$N=4$}.} 
\label{fig:4com_folded_delay}
\end{figure}

Fig.~\ref{fig:4com_folded_delay} represents the average delivery delay obtained from the folded SRN, ODE model, and the simulation when the network under-analysis consists of four communities ($N=4$), and $M$ varies from 10 to 200. As it can be seen in Fig.~\ref{fig:4com_folded_delay}, the folded model is accurate in evaluating the average delivery delay. Particularly, in Fig.~\ref{fig:4com_folded_delay}(a), as the number of nodes increases, the accuracy improves such that the percent error is less than 3\% for \mbox{$M>80$}. As it can be seen in Fig.~\ref{fig:4com_folded_delay}, the folded model is more accurate than the ODE model. According to Fig.~\ref{fig:4com_folded_delay}(a), when the number of nodes is not very large, the ODE model yields a significant error since ODE approach is rather inaccurate for networks with a moderate number of nodes due to providing limits of Markov models as mentioned in Section~\ref{sec:relWork} and \cite{zhang2007performance, yang2016delay}.
Results represented in Fig.~\ref{fig:4com_folded_delay} indicate the superiority of the folding technique in terms of accuracy compared against the ODE approach, when studying the performance of both networks with a moderate number of nodes and large-scale networks. 

\section{Scalability Analysis}\label{sec:scalability}
In this section, we investigate the scalability of the proposed monolithic and folded models and the previously presented monolithic and folded models \cite{8843121}, in terms of the number of states in the underlying Markov chains. Table~\ref{tab:state_space} represents the number of states in the underlying Markov chains of the proposed monolithic and folded models in columns \textit{Mono.} and \textit{Fold.}, respectively. As it can be seen in this table, the number of states in the underlying Markov chain of the monolithic model grows too fast as the number of communities, $N$, or the number of nodes, $M$, increases. For instance, this Markov chain, for a small network with four communities and 15 nodes ($N=4$ and $M=15$), has about 5 million states. Too much memory is needed to save this large state space, while the underlying Markov chain of the folded model for the aforementioned setting has only 16,800 states. Therefore, the folded model is scalable enough, it significantly reduces the state space.
\begin{table}	
	\renewcommand{\arraystretch}{0.9}
	\setlength{\tabcolsep}{10pt}
	\begin{center}
	\caption{Number of states in the underlying Markov chains of the proposed monolithic and folded models.}
	\label{tab:state_space}
	\end{center}
	\centering
	\begin{tabular}{c|c|c|c|c}
		\hline
		$\boldsymbol{N}$ & 
	
		 \multicolumn{2}{c|}{\bfseries 3} 
		&
		\multicolumn{2}{c}{\bfseries 4} 
		\\
		
		\hline
			 $\boldsymbol{M}$ & \bfseries Mono. & \bfseries Fold. & \bfseries Mono. & \bfseries Fold. \\

	     \hline
	     \hline
	     \bfseries 5 & 1,475  & 400 & 3,870 & 600 \\
	     \hline
	      \bfseries 10 & 56,100 & 3,300 & 287,430 & 4,950 \\
	       \hline
	      \bfseries 15 &  578,000 & 11,200  & 4,884,780  & 16,800   \\
	      \hline

	\end{tabular}
\end{table}

Table~\ref{tab:state_space} represents the number of states in underlying Markov chains of the proposed monolithic model and the previous monolithic model proposed in \cite{8843121}. As it can be observed in this table, the scalability problem of the proposed monolithic model is severer than the previous monolithic model. This is due to the fact that the proposed monolithic SRN models a more realistic network. In the network model considered in \cite{8843121}, it was assumed that each node frequently visits only one specific community, and the maximum number of local nodes in each community is $M/N$. However, the models proposed in this paper account for the possibility of moving any number of local nodes in a community. Thus, the total number of tokens in places $P^{sus\_l}_j$ and $P^{inf\_l}_j$ of submodel $Sub_j$ of the previous monolithic is at most $M/N$ tokens; while places $P^{sus\_l}_j$ and $P^{inf\_l}_j$ of submodel $Sub_{l\_j}$ of the proposed monolithic model can have totally even $M-2$ tokens, which makes the state space larger.
\begin{table}
	\renewcommand{\arraystretch}{1.1}
	\setlength{\tabcolsep}{1pt}
	\begin{center}
	\caption{Number of states in the underlying Markov chains of the proposed monolithic model and the monolithic model proposed in [1], when $N=4$.}
	\label{tab:state_space}
	\end{center}
	\centering
	\begin{tabular}{c|c|c}

		\hline
			 $\boldsymbol{M}$  & \bfseries Proposed Monolithic SRN & \bfseries Previous Monolithic SRN  \\

	     \hline
	     \hline
	     \bfseries 8 & 66,660 & 8,400  \\
	     \hline
	      \bfseries 12  & 999,570 & 192,000  \\
	       \hline
	      \bfseries 16  & 7,821,768 & 2,205,000   \\
	      \hline

	\end{tabular}
\end{table}

Table \ref{tab:comparisonOfFolded} represents the number of states in the underlying Markov chains of the folded models when the network consists of 5 communities and $M$ nodes. As it can be seen in this table, although the folded model previously published has less number of states than the folded model proposed in this paper, for the case of $M=10$, its number of states radically increases as the number of nodes increases up to five times. For example, for a network with 50 nodes ($M=50$), the underlying Markov chain of the previous folded model has about 20 million states while that of the folded model proposed herein is less than 1 million states. It is worth mentioning that the number of states of the proposed folded model even for a three times larger network ($M=150$) does not reach 20 million. Moreover, even when there are four communities and 60 nodes in the network, the previous folded model could not be solved on a system with a 64~GB memory. However, using the same system, the results reported in Figs. 12 and 13(b) are computed, which include the results obtained with the folded model for a network with four communities and 200 nodes. In conclusion, the results presented in Table \ref{tab:comparisonOfFolded} indicate that the current folded model is much more scalable than the previous one in what concerns the number of nodes ($M$). 
\begin{table}	
	\renewcommand{\arraystretch}{0.9}
	\setlength{\tabcolsep}{1pt}
	\begin{center}
	\caption{Number of states in the underlying Markov chains of the proposed folded model and the folded model proposed in \cite{8843121}, when $N=5$.}
	\label{tab:comparisonOfFolded}
	\end{center}
	\centering
	\begin{tabular}{c|c|c}

		\hline
			 $\boldsymbol{M}$ & \bfseries Proposed Folded Model & \bfseries Previous Folded Model \\

	     \hline
	     \hline
	    
	      \bfseries 10 & 6,930 & 3,552  \\
	       \hline
	      \bfseries 20 &  55,860 &  133,080     \\
	      \hline
	      \bfseries 30 &  188,790 &  1,186,080    \\
	      \hline
	      \bfseries 40 &  447,720 &  5,796,720    \\
	      \hline
	      \bfseries 50 &  874,650 &  20,211,840    \\
	      \hline

	\end{tabular}
\end{table}

\section{Conclusions and future work}\label{sec:conc}
This paper proposed two monolithic and folded SRNs to evaluate the performance of epidemic routing in MSNs. The main contribution of this paper is the performance analysis of epidemic routing, considering a network model which is more realistic than those considered in the state-of-the-art, while providing scalability. This network model is based on the skewed location visiting preferences of nodes, one of the main characteristics of MSNs. A type of first meeting time, applicable when nodes move in different areas, was analyzed. Afterwards, a monolithic SRN model was proposed to evaluate the delivery delay and the average number of transmissions by time of delivery under epidemic routing. Although the monolithic model is accurate to predict the measures of interest, it suffers from the state space explosion for networks with a large number of nodes/communities. In order to overcome this issue, we proposed an approximate model applying the folding technique to the monolithic model. This model can be used to evaluate the performance of large-scale networks without significant loss of accuracy. We also proposed an ODE model for epidemic routing and compare it with the folded model.

All the proposed models were validated against discrete-event simulation. The obtained results show that the folded model is more accurate than the ODE model. Moreover, we proved that the number of transmissions by the time of delivery follows a uniform distribution, in a general class of networks, where the positions of nodes are always i.i.d. Finally, we investigated the scalability of the proposed monolithic and folded models, contrasting with the previously presented monolithic and folded models \cite{8843121}, in terms of the number of states in the underlying Markov chains.

The current work can be extended in different ways. Other characteristics of MSNs, such as dependency of the next visited community to the currently/previously visited community and the time-dependency property of mobility \cite{Hsu:2009:MST:1665838.1665854}, can be added to the current models. It is also worth to evaluate and analyze the performance of other routing or content retrieval schemes. Moreover, using the results obtained from the proposed models, efficient routing schemes for MSNs can be developed. As discussed in Section \ref{sec:perf}, another future research direction can be further analysis of the number of transmissions when tendencies of nodes to visit a community differ.



%

\ifCLASSOPTIONcompsoc
  \section*{Acknowledgments}
\else
  \section*{Acknowledgment}
\fi

This work was partially supported by Portuguese funds through Funda\c{c}\~{a}o para a Ci\^{e}ncia e a Tecnologia (FCT) with reference UID/CEC/50021/2019.

\ifCLASSOPTIONcaptionsoff
  \newpage
\fi



%

\bibliographystyle{IEEEtran}
\bibliography{IEEEabrv,Myreferences}

\begin{thebibliography}{10}
\providecommand{\url}[1]{#1}
\csname url@samestyle\endcsname
\providecommand{\newblock}{\relax}
\providecommand{\bibinfo}[2]{#2}
\providecommand{\BIBentrySTDinterwordspacing}{\spaceskip=0pt\relax}
\providecommand{\BIBentryALTinterwordstretchfactor}{4}
\providecommand{\BIBentryALTinterwordspacing}{\spaceskip=\fontdimen2\font plus
\BIBentryALTinterwordstretchfactor\fontdimen3\font minus
  \fontdimen4\font\relax}
\providecommand{\BIBforeignlanguage}[2]{{%
\expandafter\ifx\csname l@#1\endcsname\relax
\typeout{** WARNING: IEEEtran.bst: No hyphenation pattern has been}%
\typeout{** loaded for the language `#1'. Using the pattern for}%
\typeout{** the default language instead.}%
\else
\language=\csname l@#1\endcsname
\fi
#2}}
\providecommand{\BIBdecl}{\relax}
\BIBdecl

\bibitem{8843121}
L.~{Rashidi}, A.~{Dalili-Yazdi}, R.~{Entezari-Maleki}, L.~{Sousa}, and
  A.~{Movaghar}, ``Scalable performance analysis of epidemic routing
  considering skewed location visiting preferences,'' in \emph{Proceedings of
  the IEEE 27th International Symposium on Modeling, Analysis, and Simulation
  of Computer and Telecommunication Systems (MASCOTS)}, Rennes, France, 22--24
  Oct. 2019, pp. 201--213.

\bibitem{cao2013routing}
Y.~Cao and Z.~Sun, ``Routing in delay/disruption tolerant networks: A taxonomy,
  survey and challenges,'' \emph{IEEE Communications Surveys \& Tutorials},
  vol.~15, no.~2, pp. 654--677, 2013.

\bibitem{xiao2015home}
M.~Xiao, J.~Wu, and L.~Huang, ``Home-based zero-knowledge multi-copy routing in
  mobile social networks,'' \emph{IEEE Transactions on Parallel and Distributed
  Systems}, vol.~26, no.~5, pp. 1238--1250, 2015.

\bibitem{xiao2014community}
------, ``Community-aware opportunistic routing in mobile social networks,''
  \emph{IEEE Transactions on Computers}, vol.~63, no.~7, pp. 1682--1695, 2014.

\bibitem{7961189}
G.~{Gao}, M.~{Xiao}, J.~{Wu}, K.~{Han}, L.~{Huang}, and Z.~{Zhao},
  ``Opportunistic mobile data offloading with deadline constraints,''
  \emph{IEEE Transactions on Parallel and Distributed Systems}, vol.~28,
  no.~12, pp. 3584--3599, 2017.

\bibitem{8419321}
P.~{Zuo}, Y.~{Hua}, Y.~{Sun}, X.~{Liu}, J.~{Wu}, Y.~{Guo}, W.~{Xia}, S.~{Cao},
  and D.~{Feng}, ``Bandwidth and energy efficient image sharing for situation
  awareness in disasters,'' \emph{IEEE Transactions on Parallel and Distributed
  Systems}, vol.~30, no.~1, pp. 15--28, 2019.

\bibitem{4215676}
W.-J. Hsu, T.~Spyropoulos, K.~Psounis, and A.~Helmy, ``Modeling time-variant
  user mobility in wireless mobile networks,'' in \emph{Proc. IEEE INFOCOM},
  Barcelona, Spain, 6--12 May 2007, pp. 758--766.

\bibitem{7857091}
H.~Ko, J.~Lee, and S.~Pack, ``An opportunistic push scheme for online social
  networking services in heterogeneous wireless networks,'' \emph{IEEE Trans.
  Netw. Service Manag.}, vol.~14, no.~2, pp. 416--428, 2017.

\bibitem{vahdat2000epidemic}
A.~Vahdat and D.~Becker, ``Epidemic routing for partially connected ad hoc
  networks,'' Department of Computer Science, Duke University, Durham, North
  Carolina, Tech. Rep. CS-200006, Apr. 2000.

\bibitem{zhang2007performance}
X.~Zhang, G.~Neglia, J.~Kurose, and D.~Towsley, ``Performance modeling of
  epidemic routing,'' \emph{Comput. Netw.}, vol.~51, no.~10, pp. 2867--2891,
  2007.

\bibitem{muppala1991composite}
J.~K. Muppala and K.~S. Trivedi, ``Composite performance and availability
  analysis using a hierarchy of stochastic reward nets,'' in \emph{Comput.
  Perf. Eval. Model. Tech. Tools}, G.~Balbo and G.~Serazzi, Eds.\hskip 1em plus
  0.5em minus 0.4em\relax North-Holland, The Netherlands: Elsevier Science
  Publishers B.V., 1992, pp. 335--349.

\bibitem{groenevelt2005message}
R.~Groenevelt, P.~Nain, and G.~Koole, ``The message delay in mobile ad hoc
  networks,'' \emph{Perf. Eval.}, vol.~62, no.~1, pp. 210--228, 2005.

\bibitem{IBRAHIM2007933}
M.~Ibrahim, A.~A. Hanbali, and P.~Nain, ``Delay and resource analysis in manets
  in presence of throwboxes,'' \emph{Performance Evaluation}, vol.~64, no.~9,
  pp. 933 -- 947, 2007.

\bibitem{yang2016delay}
Y.~Yang, C.~Zhao, S.~Yao, W.~Zhang, X.~Ge, and G.~Mao, ``Delay performance of
  network-coding-based epidemic routing,'' \emph{IEEE Trans. Veh. Technol.},
  vol.~65, no.~5, pp. 3676--3684, 2016.

\bibitem{ip2008performance}
Y.-K. Ip, W.-C. Lau, and O.-C. Yue, ``Performance modeling of epidemic routing
  with heterogeneous node types,'' in \emph{Proc. IEEE ICC}, Beijing, China,
  19--23 May 2008, pp. 219--224.

\bibitem{spyropoulos2009routing}
T.~Spyropoulos, T.~Turletti, and K.~Obraczka, ``Routing in delay-tolerant
  networks comprising heterogeneous node populations,'' \emph{IEEE Trans.
  Mobile Comput.}, vol.~8, no.~8, pp. 1132--1147, 2009.

\bibitem{sermpezis2016delay}
P.~Sermpezis and T.~Spyropoulos, ``Delay analysis of epidemic schemes in sparse
  and dense heterogeneous contact networks,'' \emph{IEEE Trans. Mobile
  Comput.}, vol.~16, no.~9, pp. 2464--2477, 2017.

\bibitem{picu2012forecasting}
A.~Picu and T.~Spyropoulos, ``Forecasting {DTN} performance under heterogeneous
  mobility: The case of limited replication,'' in \emph{Proc. IEEE SECON},
  Seoul, South Korea, 18--21 Jun. 2012, pp. 569--577.

\bibitem{picu2015dtn}
------, ``{DTN-M}eteo: forecasting the performance of {DTN} protocols under
  heterogeneous mobility,'' \emph{IEEE/ACM Trans. Netw.}, vol.~23, no.~2, pp.
  587--602, 2015.

\bibitem{Hsu:2009:MST:1665838.1665854}
W.-J. Hsu, T.~Spyropoulos, K.~Psounis, and A.~Helmy, ``Modeling spatial and
  temporal dependencies of user mobility in wireless mobile networks,''
  \emph{IEEE/ACM Trans. Netw.}, vol.~17, no.~5, pp. 1564--1577, 2009.

\bibitem{chaintreau2009age}
A.~Chaintreau, J.-Y. Le~Boudec, and N.~Ristanovic, ``The age of gossip: spatial
  mean field regime,'' in \emph{Proc. ACM SIGMETRICS Perf. Eval. Review}.\hskip
  1em plus 0.5em minus 0.4em\relax Seattle, WA, USA: ACM, 15--19 Jun. 2009, pp.
  109--120.

\bibitem{5719289}
J.~Whitbeck, V.~Conan, and M.~D. de~Amorim, ``Performance of opportunistic
  epidemic routing on edge-markovian dynamic graphs,'' \emph{IEEE Transactions
  on Communications}, vol.~59, no.~5, pp. 1259--1263, 2011.

\bibitem{wang2015restricted}
Q.~Wang and Q.~Wang, ``Restricted epidemic routing in multi-community delay
  tolerant networks,'' \emph{IEEE Trans. Mobile Comput.}, vol.~14, no.~8, pp.
  1686--1697, 2015.

\bibitem{8681155}
L.~{Rashidi}, R.~{Entezari-Maleki}, D.~{Chatzopoulos}, P.~{Hui}, K.~S.
  {Trivedi}, and A.~{Movaghar}, ``Performance evaluation of epidemic content
  retrieval in {DTN}s with restricted mobility,'' \emph{IEEE Transactions on
  Network and Service Management}, vol.~16, no.~2, pp. 701--714, 2019.

\bibitem{lee2013forwarding}
C.-H. Lee and D.~Y. Eun, ``On the forwarding performance under heterogeneous
  contact dynamics in mobile opportunistic networks,'' \emph{IEEE Trans. Mobile
  Comput.}, vol.~12, no.~6, pp. 1107--1119, 2013.

\bibitem{nain2005properties}
P.~Nain, D.~Towsley, B.~Liu, and Z.~Liu, ``Properties of random direction
  models,'' in \emph{Proc. IEEE INFOCOM}, 13--17 Mar. 2005, pp. 1897--1907.

\bibitem{7063247}
Q.~Xu, Z.~Su, K.~Zhang, P.~Ren, and X.~S. Shen, ``Epidemic information
  dissemination in mobile social networks with opportunistic links,''
  \emph{IEEE Trans. Emerg. Topics Comput.}, vol.~3, no.~3, pp. 399--409, 2015.

\bibitem{hsu2016enhanced}
Y.-F. Hsu and C.-L. Hu, ``Enhanced buffer management for data delivery to
  multiple destinations in {DTNs},'' \emph{IEEE Trans. Veh. Technol.}, vol.~65,
  no.~10, pp. 8735--8739, 2016.

\bibitem{lu2016distance}
Y.~Lu, W.~Wang, L.~Chen, Z.~Zhang, and A.~Huang, ``Distance-based
  energy-efficient opportunistic broadcast forwarding in mobile delay-tolerant
  networks,'' \emph{IEEE Trans. Veh. Technol.}, vol.~65, no.~7, pp. 5512--5524,
  2016.

\bibitem{Kong2008}
Z.~Kong and E.~M. Yeh, ``On the latency for information dissemination in mobile
  wireless networks,'' in \emph{Proc. ACM MobiHoc}.\hskip 1em plus 0.5em minus
  0.4em\relax ACM, 2008, pp. 139--148.

\bibitem{Zhao:2011:FRN:2030613.2030651}
S.~Zhao, L.~Fu, X.~Wang, and Q.~Zhang, ``Fundamental relationship between node
  density and delay in wireless ad hoc networks with unreliable links,'' in
  \emph{Proc. ACM MobiCom}, 2011, pp. 337--348.

\bibitem{6151313}
P.~{Basu}, S.~{Guha}, A.~{Swami}, and D.~{Towsley}, ``Percolation phenomena in
  networks under random dynamics,'' in \emph{Proceedings of the Fourth
  International Conference on Communication Systems and Networks (COMSNETS)},
  Jan 2012, pp. 1--10.

\bibitem{peres2013mobile}
Y.~Peres, A.~Sinclair, P.~Sousi, and A.~Stauffer, ``Mobile geometric graphs:
  detection, coverage and percolation,'' \emph{Probability Theory and Related
  Fields}, vol. 156, no.~1, pp. 273--305, 2013.

\bibitem{7473919}
K.~{Thilakarathna}, A.~C. {Viana}, A.~{Seneviratne}, and H.~{Petander},
  ``Design and analysis of an efficient friend-to-friend content dissemination
  system,'' \emph{IEEE Transactions on Mobile Computing}, vol.~16, no.~3, pp.
  702--715, 2017.

\bibitem{rashidi2020performance}
L.~Rashidi, D.~Towsley, A.~Mohseni-Kabir, and A.~Movaghar, ``On the performance
  analysis of epidemic routing in non-sparse delay tolerant networks,''
  \emph{arXiv preprint arXiv:2002.04834}, 2020.

\bibitem{greenwood1996guide}
P.~E. Greenwood and M.~S. Nikulin, \emph{A guide to chi-squared testing}.\hskip
  1em plus 0.5em minus 0.4em\relax John Wiley \& Sons, 1996, vol. 280.

\bibitem{peterson1981petri}
J.~L. Peterson, \emph{Petri Net Theory and the Modeling of Systems},
  1st~ed.\hskip 1em plus 0.5em minus 0.4em\relax Englewood Cliffs, NJ, USA:
  Prentice Hall, 1981.

\bibitem{ajmone1984class}
M.~Ajmone~Marsan, G.~Conte, and G.~Balbo, ``A class of generalized stochastic
  petri nets for the performance evaluation of multiprocessor systems,''
  \emph{ACM Trans. Comput. Sys.}, vol.~2, no.~2, pp. 93--122, 1984.

\bibitem{ajmone1995modelling}
M.~Ajmone~Marsan, G.~Balbo, G.~Conte, S.~Donatelli, and G.~Franceschinis,
  \emph{Modelling with Generalized Stochastic Petri Nets}, 1st~ed.\hskip 1em
  plus 0.5em minus 0.4em\relax Hoboken, NJ, USA: Wiley, 1995.

\bibitem{bause2002stochastic}
F.~Bause and P.~S. Kritzinger, \emph{Stochastic Petri Nets: An Introduction to
  the Theory}, 2nd~ed.\hskip 1em plus 0.5em minus 0.4em\relax Wiesbaden,
  Germany: Vieweg+Teubner Verlag, 2002.

\bibitem{Balazinska:2003:CMN:1066116.1066127}
M.~Balazinska and P.~Castro, ``Characterizing mobility and network usage in a
  corporate wireless local-area network,'' in \emph{Proceedings of the 1st
  International Conference on Mobile Systems, Applications and Services}, ser.
  MobiSys '03, 2003, pp. 303--316.

\bibitem{ciardo1989spnp}
G.~Ciardo, J.~Muppala, and K.~Trivedi, ``{SPNP}: stochastic petri net
  package,'' in \emph{Proc. IEEE PNPM}, Kyoto, Japan, Dec. 1989, pp. 142--151.

\bibitem{banerjee2008relays}
N.~Banerjee, M.~D. Corner, D.~Towsley, and B.~N. Levine, ``Relays, base
  stations, and meshes: enhancing mobile networks with infrastructure,'' in
  \emph{Proc. ACM MobiCom}.\hskip 1em plus 0.5em minus 0.4em\relax ACM, 14--19
  Sep. 2008, pp. 81--91.

\bibitem{hernandez2017analytical}
E.~Hernández-Orallo, M.~Murillo-Arcila, J.~C. Cano, C.~T. Calafate, J.~A.
  Conejero, and P.~Manzoni, ``An analytical model based on population processes
  to characterize data dissemination in 5{G} opportunistic networks,''
  \emph{IEEE Access}, vol.~6, pp. 1603--1615, 2018.

\end{thebibliography}

%

\begin{IEEEbiography}[{\includegraphics[width=1in,height=1.25in,clip,keepaspectratio]{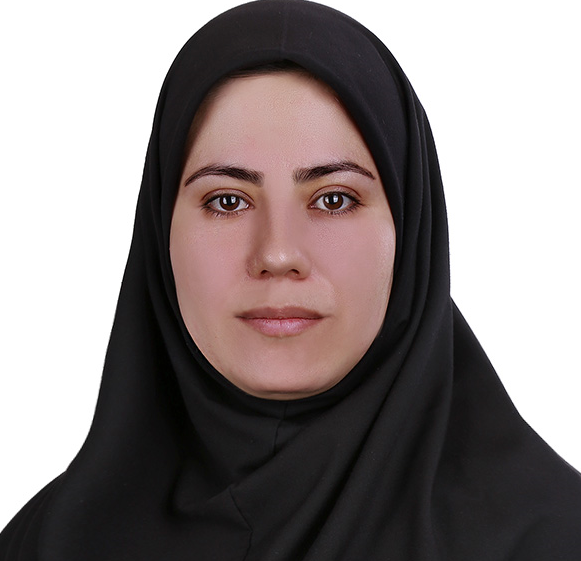}}]{Leila Rashidi} is a postdoctoral associate at the Department of Computer Science, University of Calgary, Calgary, Canada. She received her Ph.D. in Computer Engineering from the Department
of Computer Engineering, Sharif University of Technology, Iran in 2019, and the B.S. degree in Computer Engineering from the University of Tehran,
Tehran, Iran, in 2014. She was a visiting researcher at University of
Massachusetts Amherst and Imperial College London in 2017 and 2018, respectively. Her main research interests
are performance evaluation, mobile networks, optimization, and network security.
\end{IEEEbiography}

\begin{IEEEbiography}[{\includegraphics[width=1in,height=1.25in,clip,keepaspectratio]{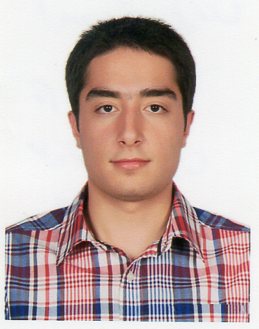}}]{Amir Dalili-Yazdi} obtained his M.Sc. degree from the Department of Computer Engineering, Sharif University of Technology, Iran in 2020.
He
received his B.Sc. in Computer Engineering, Software Field, from the Faculty of Technology and Engineering, Central Tehran Branch, Islamic Azad University, Tehran, Iran in 2017. His main research interests
are performance modeling and computer networks.
\end{IEEEbiography}

\begin{IEEEbiography}[{\includegraphics[width=0.9in,height=0.96in,clip,keepaspectratio]{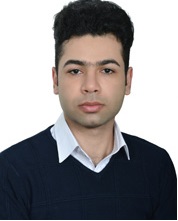}}]{Reza Entezari-Maleki}
received the B.S. and M.S. degrees from the Iran University of Science and Technology, Tehran, Iran, in 2007 and 2009, respectively, and the Ph.D. degree from the Sharif University of Technology, Tehran, Iran, in 2014, all in computer engineering. He worked as a post-doctoral researcher at the School of Computer Science, Institute for Research in Fundamental Sciences (IPM), Tehran, Iran, from 2015 to 2018. He is currently an assistant professor in Iran University of Science and Technology. His main research interests are performance/dependability modeling and evaluation in distributed computing systems.
\end{IEEEbiography}

\begin{IEEEbiography}[{\includegraphics[width=0.9in,height=0.96in,clip,keepaspectratio]{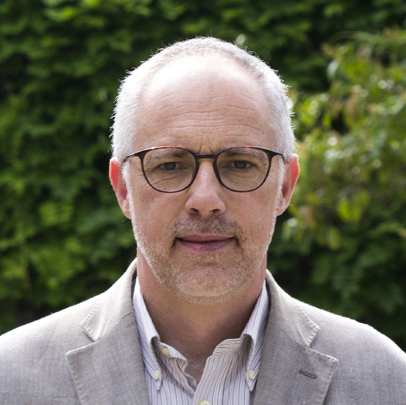}}]{Leonel Sousa}
received the PhD degree in electrical and computer engineering from the Instituto Superior Tecnico (IST), Universidade de Lisboa (UL), Lisbon, Portugal, in 1996. Since 1996, he has been with IST, where he is currently the chair of the Department of Electrical and Computer Engineering, and has been a full professor since 2010. In 2016, he was a visiting professor with Tsukuba University, Tsukuba, Japan, with a JSPS Invitation Fellowship for Research in Japan, and Carnegie Mellon University, Pittsburgh, PA. His research interests include computer architectures, high performance computing, and multimedia systems. He has contributed more than 250 papers for international journals and conferences and to the organization of several international conferences. He is currently an associate editor of the IEEE Transactions on Computers. He is also a distinguished scientist of the ACM.
\end{IEEEbiography}

\begin{IEEEbiography}[{\includegraphics[width=1in,height=1.25in,clip,keepaspectratio]{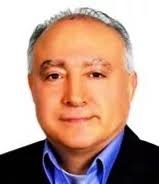}}]{Ali Movaghar} is a Professor in the Department of Computer Engineering at Sharif University of Technology. He received his B.S. degree in Electrical Engineering from the University of Tehran in 1977, and M.S. and Ph.D. degrees in Computer, Information, and Control Engineering from the University of Michigan, in 1979 and 1985, respectively.
He visited the Institut National de Recherche en Informatique et en Automatique in Paris, France and the Department of Electrical Engineering and Computer Science at the University of California, Irvine in 1984 and 2011, respectively, worked at AT\&T Information Systems in Naperville, IL in 1985-1986, and taught at the University of Michigan, Ann Arbor in 1987-1989.
His research interests include performance/dependability modeling and formal verification of wireless networks and distributed real-time systems.
\end{IEEEbiography}




\end{document}